\documentclass[journal,twoside,web]{ieeecolor}
\pdfoutput=1
\usepackage{tmi}
\usepackage{cite}
\usepackage{amsmath,amssymb,amsfonts}
\usepackage{algorithmic}
\usepackage{graphicx}
\usepackage{textcomp}
\usepackage{booktabs}
\usepackage{multirow}
\usepackage{caption}
\usepackage{subcaption}
\usepackage[capitalise]{cleveref}
\usepackage{datatool}
\usepackage[acronym,shortcuts]{glossaries}
\def\BibTeX{{\rm B\kern-.05em{\sc i\kern-.025em b}\kern-.08em
    T\kern-.1667em\lower.7ex\hbox{E}\kern-.125emX}}
\graphicspath{ {figures/} }

\DeclareMathOperator*{\argmax}{\arg\!\max}

\newacronym{dl}{DL}{deep learning}
\newacronym{em}{EM}{Expectation-Maximization}
\newacronym{cnn}{CNN}{Convolutional Neural Network}
\newacronym{crf}{CRF}{Conditional Random Field}
\newacronym{mrf}{MRF}{Markov Random Field}
\newacronym{fcn}{FCN}{Fully Convolutional Networks}
\newacronym{rcnn}{RCNN}{Recurrent \gls*{cnn}}
\newacronym{roi}{ROI}{Region of Interests}
\newacronym{ct}{CT}{Computed Tomography}
\newacronym{mr}{MR}{Magnetic Resonance}
\newacronym{tse}{TSE}{turbo spin echo}
\newacronym{xent}{xent}{crossentropy}
\newacronym{ftl}{FTL}{Focal Tversky Loss}
\newacronym{mae}{MAE}{Mean Absolute Error}
\newacronym{iou}{IOU}{intersection-over-union}
\newacronym{msd}{MSD2018}{Medical Segmentation Decathlon 2018}
\newacronym{imaf}{IMAF}{intermuscular adipose fat}
\newacronym{scf}{SCF}{subcutaneous fat}

\begin{document}
\title{Training CNN Classifiers for Semantic Segmentation using Partially Annotated Images: with Application on Human Thigh and Calf MRI}

\author{Chun Kit Wong, Stephanie Marchesseau, Maria Kalimeri, Tiang Siew Yap, Serena S. H. Teo, Lingaraj Krishna, Alfredo Franco-Obreg\'{o}n, Stacey K. H. Tay, Chin Meng Khoo, Philip T. H. Lee, Melvin K. S. Leow, John J. Totman, and Mary C. Stephenson
\thanks{This work was partially funded by the NUHS Imaging Core Grant [NMRC/CG/M009/2017\_NUH/NUHS]. MRI acquisition of our calf-tissue dataset was funded by Nestl\'{e} Institute of Health Sciences (NIHS) and the EpiGen Consortium. (\textit{Corresponding author: Chun Kit Wong}.)}
\thanks{Chun Kit Wong, Tiang Siew Yap, Serena S. H. Teo, and Mary C. Stephenson are with the Clinical Imaging Research Centre, Yong Loo Lin School of Medicine, National University of Singapore (email: wongck@nus.edu.sg).}
\thanks{Stephanie Marchesseau was with the Clinical Imaging Research Centre, Yong Loo Lin School of Medicine, National University of Singapore. She is now with Savana Medica, Madrid, Spain.}
\thanks{Maria Kalimeri and John J. Totman were with the Clinical Imaging Research Centre, Yong Loo Lin School of Medicine, National University of Singapore.}
\thanks{Lingaraj Krishna is with the Department of Orthopaedic Surgery, National University Hospital, National University Health System, Singapore.}
\thanks{Alfredo Franco-Obreg\'{o}n is with (1) Department of Surgery, Yong Loo Lin School of Medicine, National University of Singapore, (2) BioIonic Currents Electromagnetic Pulsing Systems Laboratory, BICEPS, National University of Singapore, and (3) Institute for Health Innovation \& Technology, iHealthtech, National University of Singapore}
\thanks{Stacey K. H. Tay is with (1) Department of Paediatrics, Yong Loo Lin School of Medicine, National University of Singapore, and (2) KTP-National University Children’s Medical Institute, National University Health System, Singapore}
\thanks{Chin Meng Khoo is with (1) Department of Medicine, Yong Loo Lin School of Medicine, National University of Singapore, and (2) Department of Medicine, National University Health System, Singapore}
\thanks{Philip T. H. Lee is with the Agency for Science, Technology, and Research, Singapore}
\thanks{Melvin K. S. Leow is with (1) Clinical Nutrition Research Centre, Singapore Institute for Clinical Sciences, Agency for Science, Technology, and Research, Singapore, (2) Cardiovascular and Metabolic Disorders Program, Duke-National University of Singapore Medical School, Singapore, (3) Lee Kong Chian School of Medicine, Nanyang Technological University, Singapore, and (4) Department of Endocrinology, Tan Tock Seng Hospital, Singapore}
}
\maketitle
\begin{abstract}
\textit{Objective}: Medical image datasets with pixel-level labels tend to have a limited number of organ or tissue label classes annotated, even when the images have wide anatomical coverage. With supervised learning, multiple classifiers are usually needed given these partially annotated datasets. In this work, we propose a set of strategies to train one single classifier in segmenting all label classes that are heterogeneously annotated across multiple datasets without moving into semi-supervised learning.
\textit{Methods}: Masks were first created from each label image through a process we termed presence masking. Three presence masking modes were evaluated, differing mainly in weightage assigned to the annotated and unannotated classes. These masks were then applied to the loss function during training to remove the influence of unannotated classes.
\textit{Results}: Evaluation against publicly available CT datasets shows that presence masking is a viable method for training class-generic classifiers. Our class-generic classifier can perform as well as multiple class-specific classifiers combined, while the training duration is similar to that required for one class-specific classifier. Furthermore, the class-generic classifier can outperform the class-specific classifiers when trained on smaller datasets. Finally, consistent results are observed from evaluations against human thigh and calf MRI datasets collected in-house.
\textit{Conclusion}: The evaluation outcomes show that presence masking is capable of significantly improving both training and inference efficiency across imaging modalities and anatomical regions. Improved performance may even be observed on small datasets.
\textit{Significance}: Presence masking strategies can reduce the computational resources and costs involved in manual medical image annotations.
\end{abstract}


\section{Introduction}
\label{sec:introduction}
Medical image analysis has recently been dominated by \gls*{dl} methods \cite{litjens_deeplearningsurvey}. This is especially true for semantic segmentation, where most state-of-the-art algorithms involve the use of \gls*{cnn}. For instance, the U-Net architecture \cite{ronneberger2015_unet} has been widely used in semantic segmentation to generate 2D pixel-level label map. Numerous extensions of the architecture have since been proposed, such as the 3D U-Net \cite{cicek2016_3Dunet} and V-Net \cite{milletari2016_vnet}. %, or more recently the R2U-Net \cite{alom_r2unet} which combines ResNet \cite{he_resnet} and \gls*{rcnn} with U-Net.

While \gls*{dl} has been successful in extracting highly implicit pattern from the data, performance in its biomedical application is often limited by availability of annotated data itself \cite{litjens_deeplearningsurvey}. Specifically in semantic segmentation, datasets manually annotated with pixel-level label map are often small, due to intense labour and high technical expertise requirement on the annotator. This has lead to development of alternative annotation and training strategies, many of which extend concepts from weakly- and semi-supervised learning \cite{cheplygina_semisupervisedsurvey} to train classifiers using both labelled and unlabelled images \cite{zhu_introtosemisupervised}. Weak labels such as point, bounding-box, or image level labels \cite{papandreou2015_weaksemisupervised,bearman_whatsthepoint,rajchl2016_bboxsegmentation} may also be utilized in addition to the strong, pixel-level labels. One specific implementation, commonly known as self-supervised learning, uses the \gls*{em} algorithm \cite{dempster_em} to alternate between predicting strong labels for weakly- and un-labelled feature images, and optimizing classifiers using both manually-annotated and the predicted strong labels as ground-truth \cite{cheplygina_semisupervisedsurvey}. Since weakly- and un-labelled biomedical images are often much easier to collect \cite{lin_microsoftcoco}, larger dataset can be obtained with a reduced labour budget. However, such approaches come with their own drawbacks \cite{fabio_semisupervised_drawback,aarti_unlabeleddata_drawback}, and may not work better than the fully supervised approach.

%Since weak labels \cite{lin_microsoftcoco} and unlabelled data are often much easier to be collected, larger training dataset can be obtained with the same labour budget, which in turn leads to wide adoption of semi-supervised learning in biomedical applications \cite{cheplygina_semisupervisedsurvey}. However, semi-supervised learning has its own drawbacks and does not always work better than the supervised approach \cite{fabio_semisupervised_drawback,aarti_unlabeleddata_drawback}. For simplicity, exploration of the proposed method under semi-supervised setting is hence omitted from this work.

Recently, a number of works on multi-organ analysis have been published \cite{cerrolaza2019_multiorgan}, with many taking \gls*{dl} approaches, as expected \cite{shin2013_multiorgan1,roth2015_multiorgan2,yan2016_multiorgan3,zhou2016_majorityvoting,wang2018_multiorgan4}. In such contexts, there exists an additional form of the data scarcity problem that tends to get neglected. Multi-organ semantic segmentation datasets, such as the 20-class VISCERAL \cite{toro2016_visceral} or the 19-class "Computational Anatomy" \cite{zhou2016_majorityvoting} datasets, remain relatively rare as compared to single-organ datasets. More typically, only a limited number of organs are annotated, even when the acquired feature images may cover multiple organs. With only single-organ datasets available, multiple classifiers have to be trained to segment all organs of interest if a fully supervised learning setting is desired. This potentially leads to inefficiencies, especially among label classes with similar appearance, since the classifiers have to separately learn a redundant set of features. Furthermore, to obtain predictions for all label classes, multiple inferences have to be made separately with the classifiers, which is less optimal especially for inference-on-the-edge applications.

In this work, we propose a method to train one multi-organ classifier utilizing single-organ datasets under supervised learning. This is done by first merging the single-organ datasets (with proper class label mapping function) into a larger, partially labelled multi-organ dataset. We then developed a few strategies to prevent un-annotated label classes from contributing to the loss function, which we term collectively as \textit{presence masking}. These strategies are evaluated using openly available \gls*{ct} datasets from the \gls*{msd} \cite{msd2018}. Here, we showed that a multi-organ classifier trained using our method is capable of delivering performance on-par with single-organ classifiers. Additional evaluations are performed under small training dataset scenario, simulated by shrinking the \gls*{msd} dataset, with results indicating that the multi-organ classifier actually outperforms single-organ classifiers among label classes with similar appearance.

To further verify performance of our proposed method, we applied the presence masking strategies on human thigh and calf \gls{mr} images. Much like other applications, annotated label maps of human thigh and calf are often needed for quantitative musculoskeletal analysis \cite{ziegenfuss2002_muscleAthleteCreatine, wroblewski2011_muscleAthlete, finanger2012_muscleDMD, hiba2012_muscleMD}. Since annotated thigh and calf datasets with open access are rarely available, most publications resorted to dedicatedly annotating their own images for classifier training \cite{ghosh2017_thighsegmentation,ezak2018_thighsegmentation}. Here, we show that it is equivalent to using partially labelled images from various sources, even if they're heterogeneous in many way (healthy/pathologic subjects, imaging protocol, label class annotated, etcs).

The rest of this paper is organized as follows: \Cref{sec:methodology} provides detailed descriptions of the methodology. In particular, \cref{sec:partiallylabel} explains our proposed presence masking strategies to handle the situation whereby an organ/tissue potentially exists on the feature image but is missing from the label map. \Cref{sec:exp_evaluations} describes technical details behind the evaluation experiments. In total, three experiments are performed, with the corresponding results shown and discussed under \cref{sec:results_and_discussions}. \Cref{sec:expconfigs_specificvsgeneric,sec:result_specificvsgeneric} compares performance of the proposed multi-organ classifiers with single-organ classifiers in segmenting the \gls*{msd} dataset. \Cref{sec:expconfigs_smalldataset,sec:result_smalldataset} further evaluate the multi-organ classifiers' performance with small training datasets, simulated by shrinking the original training dataset into smaller proportions. \Cref{sec:expconfigs_musc,sec:result_musc} applies the validated method on segmenting \gls{mr} images of the human thigh and calf. \Cref{sec:potential_improvements} discusses about potential integrations of the proposed presence masking strategies with existing \gls{dl} methodologies.

\section{Methodology}
\label{sec:methodology}

\begin{figure*}[!t]
	\centering
	\includegraphics[width=\linewidth]{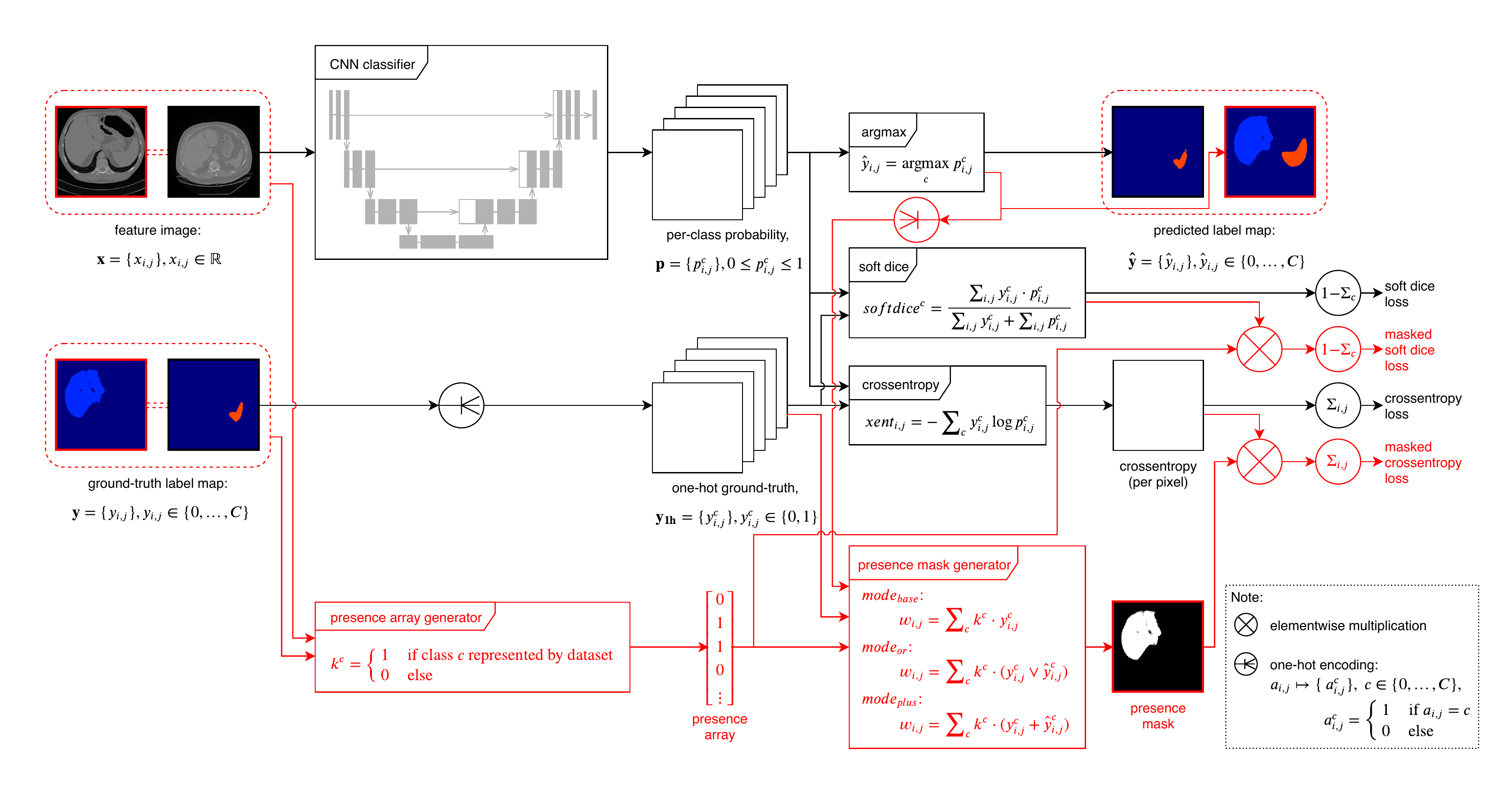}
	\caption{Workflow in training \gls*{cnn} classifier for semantic segmentation. Black components indicate steps taken when training images are completely segmented, while red components indicate additional steps when they are partially segmented. Note that $(i,j)$ is used to represent pixel location in the 2D images, in place of the dimension-independent $m$ used in the main text.}
	% Back-propagation of the gradients from loss functions to the \gls*{cnn} classifier is omitted for readability. }
	\label{fig:workflow}
\end{figure*}

\subsection{Notation}
\label{sec:notation}
\gls*{ct} and \gls*{mr} feature image is denoted $\mathbf{x}=\{x_{m}\},m\in\{1,\dots,M\}$, with $x_{m}$ being the pixel value at location $m$. The corresponding ground-truth label map is denoted $\mathbf{y}=\{y_{m}\}$, with $y_{m}=c,c\in\{0,\dots,C\}$ indicates that $x_{m}$'s ground-truth class is $c$, which can either be the background class $0$ or any of the $C$ foreground classes.

\gls*{cnn} classifier is denoted $f(\mathbf{x},\boldsymbol{\theta})$, with $\boldsymbol{\theta}$ parametrizing all model parameters. It is assumed that the final layer of the classifier is the softmax function. Pixel-wise, per-class probability that $x_{m}$ belongs to class $c$ is denoted $\mathbf{p}=\{p_{m}^{c}\}$, with the corresponding predicted label map denoted as $\mathbf{\hat{y}}=\{\hat{y}_{m}\}$. For both $y_{m}$ and $\hat{y}_{m}$, the $c$-superscripted counterpart denotes the label map in one-hot representation:
\begin{equation}
y_{m}^{c} = \begin{cases}
1 & \text{if } y_{m} = c\\
0 & \text{else}
\end{cases}
\label{eq:onehot}
\end{equation}

%Additionally, a binary, dataset-level presence array $\mathbf{k}=\{k_{c}\}$ is defined such that when $k_{c}=true$, $y_{m}=c$ if and only if $x_{m}$ shall be classified into class $c$. When $k_{c}=false$, $y_{m}$ can be of any value if $x_{m}$ shall be classified into class $c$. In another words, $k_{c}$ encodes whether or not the value $c$ in label map $y$ is valid. Locations with $y_{m}=c$ shall not be ignored only if $k_{c}=true$.

A dataset $D=\{(\mathbf{x},\mathbf{y})_{n}\},n\in\{1,\dots,N\}$ with $N$ image pairs $(\mathbf{x},\mathbf{y})$ is completely labelled when each of the $C$ foreground classes are fully annotated on $\mathbf{y}$ whenever it appears on $\mathbf{x}$. Formally, this implies that $y_{m}=c$ if and only if $x_{m}$ should be assigned the ground-truth label class $c$, for every class $c\in\{0,\dots,C\}$. Conversely, in a partially labelled dataset, $y_{m}\neq c$ does not necessarily indicate that $x_{m}$ should not be assigned to class $c$. For example, when a liver dataset is combined with a spleen dataset (with proper mapping such that $c=1$ represents liver while $c=2$ represents spleen), the resultant dataset is partially labelled. Images originating from the liver dataset may contain pixels $x_{m}$ corresponding to spleen, but are assigned the background class label (i.e. $y_{m}=0$) since spleen is not annotated in its parent dataset. In another words, $c\in\{0,2\}$ shall be ignored in $\mathbf{y}$.

%Classifiers trained with completely or partially labelled dataset are then designated as specific- or generic-classifiers respectively.

%A completely-labelled dataset is then defined as the set of $N$ feature-label image pairs $D=\{(\mathbf{x},\mathbf{y})_{n}\}, n\in\{1,\dots,N\}$ with $k_{c}=true$ for all $c\in\{0,\dots,C\}$. If $k_{c}=false$ for any class $c$, the dataset is considered partially-labelled.

\subsection{Presence masking for partially labelled dataset}
\label{sec:partiallylabel}

\begin{figure*}[!htb]%[!ht]
	\centering
	\begin{subfigure}[t]{0.9\linewidth}
		\centering
		\includegraphics[width=\linewidth,page=1]{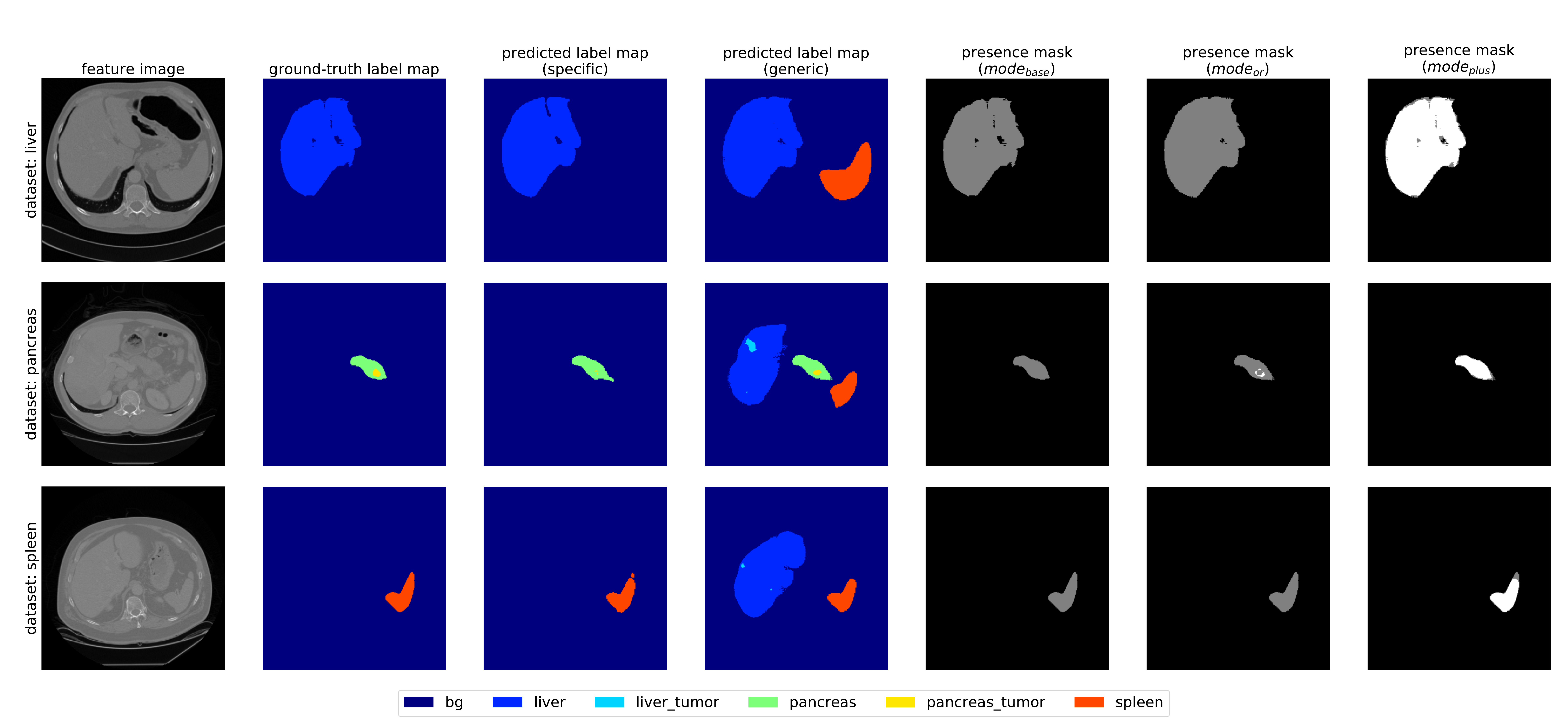}
		\caption{}
		\label{fig:screenshot_and_presence_masks_msd2018}
	\end{subfigure}
	\begin{subfigure}[t]{0.9\linewidth}
		\centering
		\includegraphics[width=\linewidth,page=2]{MUSC/screenshots}
		\caption{} 
		\label{fig:screenshot_and_presence_masks_calf}
	\end{subfigure}
	\begin{subfigure}[t]{0.9\linewidth}
		\centering
		\includegraphics[width=\linewidth,page=4]{MUSC/screenshots}
		\caption{}
		\label{fig:screenshot_and_presence_masks_thigh}
	\end{subfigure}
	\caption{Screenshots of feature image and label map from (\protect\subref{fig:screenshot_and_presence_masks_msd2018}) \acs{msd}, (\protect\subref{fig:screenshot_and_presence_masks_calf}) calf, and (\protect\subref{fig:screenshot_and_presence_masks_thigh}) thigh datasets, together with example predictions by the specific- and generic-classifiers (see \cref{sec:exp_evaluations}). Presence mask in different modes are generated using the ground-truth label maps and the corresponding generic-classifier's predictions. Note how the generic-classifiers predict all label classes present in the entire (partially labelled) dataset. Generic-classifier's prediction for the thigh dataset is omitted since it is completely labelled.}
	\label{fig:screenshot_and_presence_masks}
\end{figure*}

%Dataset $D=\{(\mathbf{x},\mathbf{y})_{n}\}, n\in\{1,\dots,N\}$ is defined to be completely labelled when $y_{m}=c$ if and only if $x_{m}$ should be assigned the ground-truth label class $c$, for every class $c\in\{0,\dots,C\}$. Conversely, in a partially labelled dataset, $y_{m}\neq c$ does not necessarily indicate that $x_{m}$ should not be assigned to class $c$. For example, when a liver dataset is combined with a spleen dataset (with proper mapping such that $c=1$ represents liver while $c=2$ represents spleen), the resultant dataset is partially labelled. Images originate from the liver dataset may contain pixels $x_{m}$ corresponding to spleen, but are assigned $y_{m}=0$ (i.e. the background class label) since spleen is not labelled in its parent dataset. In another words, $c\in\{0,2\}$ shall be ignored in $\mathbf{y}$.

\cref{fig:workflow} depicts the workflow in supervised semantic segmentation. An overview of supervised semantic segmentation can be found under \cref{sec:method_overview} in the supplementary material. Independent of the choice of loss function, the workflow described under \cref{sec:method_overview} works only when the training dataset is completely labelled. With partially labelled training dataset, classes not annotated should be masked out while computing the loss function $J(\theta)$. This leads to our proposed presence masking strategies.

\subsubsection{Presence masking the dice losses}
To begin with, a dataset-level presence array $\mathbf{k}=\{k^{c}\}, k^{c}=\{0,1\}$ is defined for each $(\mathbf{x},\mathbf{y})$ pairs, with $k^{c} = 0$ indicating that class $c$ is not annotated manually in the dataset. In another words, any presence of class $c$ in $\mathbf{y}$ is invalidated. This can then be directly applied onto the class-aggregation term in the dice losses (see \cref{eq:softdiceloss,eq:logdiceloss} in the supplementary material) to obtain the masked soft dice loss:
\begin{equation}
J_{dice_{soft}}(\boldsymbol{\theta}) = \frac{\sum_{c}^{C}\nolimits k^{c}\cdot (1 - softdice^{c})}{\sum_{c}^{C}\nolimits k^{c}}
\label{eq:masked_softdiceloss}
\end{equation}
or the masked log dice loss:
\begin{equation}
J_{dice_{log}}(\boldsymbol{\theta}) = \frac{\sum_{c}^{C}\nolimits k^{c}\cdot (-\log{softdice^{c}})}{\sum_{c}^{C}\nolimits k^{c}}
\label{eq:masked_logdiceloss}
\end{equation}

\subsubsection{Presence masking the crossentropy loss}
Crossentropy loss, however, requires extra steps. For class $c$ where $k^{c}=true$, false positive errors occur at location $m$ where $y_{m}\neq c$ and $\hat{y}_{m}=c$ hold. When $y_{m}\neq c$, $k^{c}\cdot y_{m}^{c}=0$ regardless of $k^{c}$. This means that false positive errors are potentially ignored if $\mathbf{k}$ is applied directly to the class aggregation term in the crossentropy loss. To mitigate this, a presence mask $\mathbf{w}=\{w_{m}\}, m\in\{1,\dots,M\}$ can be defined and applied to the pixel aggregation term instead:
\begin{equation}
J_{xent}(\boldsymbol{\theta}) = - \sum_{m}^{M}\nolimits w_{m}\sum_{c}^{C}\nolimits y_{m}^c \log p_{m}^c
\label{eq:masked_xentloss}
\end{equation}
% We propose multiple ways to construct $\mathbf{w}$. The most naive approach, which we denote $mode_{base}$, would be to set $w_{m}=1$ at every location $m$ where $y_{m}=c$ and $k^{c}=1$, or $w_{m}=0$ whenever $y_{m}=c$ and $k^{c}=0$. This can be formulated compactly as:
% \begin{equation}
% w_{m} = \sum_{c}^{C}\nolimits k^{c} \cdot y_{m}^{c}
% \label{eq:mode_base}
% \end{equation}
% However, $mode_{base}$ also ignores false positive errors when $\mathbf{y}$ is partially labelled. To resolve that, $\mathbf{\hat{y}}$ has to be incorporated while constructing $\mathbf{w}$. One possible way would be to perform elementwise-or between $\mathbf{y}$ and $\mathbf{\hat{y}}$ ($mode_{or}$):
% \begin{equation}
% w_{m} = \sum_{c}^{C}\nolimits k^{c} \cdot (y_{m}^{c} \lor \hat{y}_{m}^{c})
% \label{eq:mode_or}
% \end{equation}
% or by summing them up ($mode_{plus}$):
% \begin{equation}
% w_{m} = \sum_{c}^{C}\nolimits k^{c} \cdot (y_{m}^{c} + \hat{y}_{m}^{c})
% \label{eq:mode_plus}
% \end{equation}
% Essentially, the various modes differ in terms of how the true positive and false positive/negative pixels are weighted (see \cref{fig:screenshot_and_presence_masks}). Mode selection can then be tuned as hyperparameter during classifier training.
We propose multiple way, which we denote as mode, to construct $\mathbf{w}$:%, which we denote as the presence masking modes:
\begin{itemize}
    \item $mode_{base}$: set $w_{m}=1$ at every location $m$ where $y_{m}=c$ and $k^{c}=1$, and $w_{m}=0$ wherever $y_{m}=c$ and $k^{c}=0$. This can be formulated compactly as:
        \begin{equation}
        w_{m} = \sum_{c}^{C}\nolimits k^{c} \cdot y_{m}^{c}
        \label{eq:mode_base}
        \end{equation}
    
    \item $mode_{or}$: same as $mode_{base}$, but extended to also include every location $m$ where $\hat{y}_{m}=c$. In other words, take the elementwise-or between $\mathbf{y}$ and $\mathbf{\hat{y}}$, instead of $\mathbf{y}$ alone:
        \begin{equation}
        w_{m} = \sum_{c}^{C}\nolimits k^{c} \cdot (y_{m}^{c} \lor \hat{y}_{m}^{c})
        \label{eq:mode_or}
        \end{equation}
    
    \item $mode_{plus}$: same as $mode_{or}$, but take the elementwise-sum between $\mathbf{y}$ and $\mathbf{\hat{y}}$ instead of elementwise-or:
        \begin{equation}
        w_{m} = \sum_{c}^{C}\nolimits k^{c} \cdot (y_{m}^{c} + \hat{y}_{m}^{c})
        \label{eq:mode_plus}
        \end{equation}
\end{itemize}

Effectively, $mode_{base}$ is equivalent to generating the presence mask $\mathbf{w}$ by binary thresholding the ground truth label map $\mathbf{y}$, such that only foreground class $c$ with $k^{c}=1$ in the presence array $\mathbf{k}$ are mapped to 1 in $\mathbf{w}$. However, this naive approach does not resolve the issue mentioned at the beginning of this section, since it also ignores false positive errors when $\mathbf{y}$ is partially labelled. To avoid masking out false positive errors, the predicted label map $\mathbf{\hat{y}}$ has to be incorporated while constructing $\mathbf{w}$. Among many possible approaches, this could be achieved by computing either the elementwise-or ($mode_{or}$) or elementwise-sum ($mode_{plus}$) between $\mathbf{y}$ and $\mathbf{\hat{y}}$. Essentially, these two modes only differ in terms of how the true positive and false positive/negative pixels are weighted (see \cref{fig:screenshot_and_presence_masks}): $mode_{or}$ assigns equal weightage to all these pixels, while $mode_{plus}$ assigns heavier weightage to the true positive pixels.

%(Note the difference between this and the image-level annotations from weakly-supervised segmentation, where a positive image-level annotation for class $c$ indicates that $y_{m}=c$ can be observed for at least one location $m$ \cite{papandreou2015_weaksemisupervised}.)

To avoid any confusion, note the difference between presence array $\mathbf{k}$ and the image-level annotation commonly used in a weakly-supervised setting \cite{papandreou2015_weaksemisupervised}. Presence array indicates whether or not a label class $c$ is represented in the dataset. When $k_{c}=true$, even when $y_{m} \neq c$ for all $m$, it indicates a true negative situation. Any prediction that says $\hat{y}_m=c$ should be treated as a false positive error. On the other hand, a positive image-level annotation for class $c$ indicates that if the pixel-level label map $\mathbf{y}$ exists, $y_{m}=c$ can be observed for at least one location $m$.

\subsubsection{Presence masking and completely labelled image}
\label{sec:presence_masking_complete_data}
While our presence masking strategies are developed to handle partially labelled images, it is also applicable to completely labelled images, which is essentially the special case where $k_{c}=true$ for all label class $c$. Under this special case, the masked dice losses (\cref{eq:masked_softdiceloss,eq:masked_logdiceloss}) simply become the unmasked versions scaled by a constant value $1/\sum_c^C k^c$. Crossentropy loss masked with $mode_{base}$ presence mask also converges back to the unmasked version. Presence mask under $mode_{or}$ and $mode_{plus}$, however, can then be seen as an attempt to improve the loss functions, by weighting each pixel differently. Hence, this should be taken into consideration while performing any performance evaluations (see \cref{sec:expconfigs_specificvsgeneric,sec:result_specificvsgeneric}).

\section{Experimental evaluations}
\label{sec:exp_evaluations}

To facilitate discussion, we denote classifiers trained with completely labelled single-organ dataset as \textit{specific}-classifiers. Alternatively, those trained with partially labelled multi-organ dataset are denoted \textit{generic}-classifiers (see \cref{fig:screenshot_and_presence_masks}). Classifiers are also addressed based on the loss function and mode of presence mask (where included). For example, a $xent_{or} + 0.1 * dice_{soft}$ classifier refers to the classifier trained with loss function $J_{xent}(\boldsymbol{\theta}) + 0.1 \cdot J_{dice_{soft}}(\boldsymbol{\theta})$, where the presence mask applied pixel-wise to the crossentropy loss is in $mode_{or}$.

\subsection{Datasets}
\label{sec:datasets}
All evaluations in this work leverage on annotated image pairs $(\mathbf{x},\mathbf{y})$ from two main sources. For reproducibility, the proposed method under \cref{sec:methodology} was evaluated with image pairs $(\mathbf{x},\mathbf{y})$ from the \gls*{msd}; specifically the liver (131 pairs), pancreas (281 pairs), and spleen (41 pairs) datasets amongst the 10 available datasets. All 3 datasets contain \gls*{ct} feature images of the abdominal region. In addition to healthy tissues, tumors were also annotated in both liver and pancreas datasets. Images from these datasets were merged into one large partially-labelled dataset with proper class-mapping scheme.

%To investigate the proposed method's versatility and practical applications, it is further evaluated against in-house datasets with \gls*{mr} feature images of human thigh and calf.
\gls*{mr} feature images of human thigh and calf were acquired at the Clinical Imaging Research Centre, National University of Singapore. These \gls*{mr} images were acquired using \gls*{tse} sequence and Siemens Magnetom 3T Prisma or TrioTim systems. Ground-truth label maps were manually annotated by certified radiographers using MITK (v2015.05.02) \cite{WOLF2005594_mitk}. The calf dataset is partially labelled, and comprised of two sub-datasets: a calf-muscle dataset containing 22 left calves acquired from healthy subjects, with 6 muscle groups annotated, and a calf-tissue dataset containing 84 left and right calves acquired simultaneously from 42 subjects, with 3 tissue classes (bone, blood vessels, \gls*{scf}) annotated. On the other hand, the thigh dataset is comprised of completely labelled $(\mathbf{x},\mathbf{y})$ pairs with 4 tissue classes (bone, blood vessels, \gls*{scf}, and \gls*{imaf}) and 11 muscle group classes annotated, with 40 thighs acquired from 20 double-visit surgical subjects. Full detail of the datasets can be found in the supplemented \cref{tab:musc_dataset_desc}.

\subsection{Pre- and post-processing}
\label{sec:pre_post_processing}

For the train-test cycle, all images from the \gls*{msd} dataset were resampled with bilinear interpolation to match the standardized $1.5x1.5\text{mm}^2$ in-plane resolution and $3.5\text{mm}$ slice thickness. Thigh images were resampled to $0.78x0.78\text{mm}^2$ in-plane resolution, with no resampling in the slice direction. These were then cropped/zero-padded to form $256x256$ slices. On the other hand, images from the calf-tissue dataset were cropped into single calves and combined with the calf-muscle dataset, before all images were resampled to $0.54x0.54\text{mm}^2$ in-plane resolution and cropped/zero-padded to form $320x320$ slices.

At inference time, feature images $\mathbf{x}$ were similarly resampled to the standardized resolution. Predicted label maps $\mathbf{\hat{y}}$ from the trained classifiers were then resampled (with nearest-neighbour interpolation) to match the original resolution and array size. For all evaluation results reported in the following sections, dice scores were calculated between $\mathbf{y}$ and $\mathbf{\hat{y}}$ in their original resolution and array size.

Note that the images were resampled aggressively to expedite the train-test cycle, as well as to reduce computational cost needed. However, this had also inevitably led to interpolation error on the final result. Performance is expected to improve when all images are processed in their original resolution.

\subsection{Training details}
\label{sec:training_details}
Neural networks were implemented with Tensorflow (v1.4) in Python. All \gls*{cnn} classifiers used in this work were instances of the 2D U-Net model \cite{ronneberger2015_unet} with additional batch normalization layers after every convolution and deconvolution layers to improve robustness against variance within each mini-batch \cite{baumgartner2017}. % Source code used for evaluations will be released online together with paper acceptance.

Much of the training hyperparameters were empirically fixed. Network parameters $\boldsymbol{\theta}$ were initialized with Xavier initialization \cite{glorot2010_xavierinit} and optimized using Adam optimizer \cite{KingmaB14_AdamOptimizer}. A default learning rate of $10^{-4}$ was chosen, and was reduced to $10^{-5}$ only if the training failed to progress. Patch-based approaches were omitted for simplicity. All classifiers were trained until convergence in a cluster equipped with Intel E5-2690v3 and NVIDIA Tesla K40.

Classifiers were trained using mini-batches of $(\mathbf{x},\mathbf{y})$ from $D_{train}$, with 15 image pairs per batch. To improve training stability of the generic-classifiers, these image pairs were drawn from their parent dataset using stratified sampling. Each batch was a mixture of $(\mathbf{x},\mathbf{y})$ from each parent dataset at a constant proportion. Then, for both specific- and generic-classifiers, half of the image pairs were labelled with at least one foreground class, while the other half did not contain any foreground labels.

\subsection{Experiment configurations}
\label{sec:experiment_configs}

\subsubsection{Specific- versus generic-classifiers}
\label{sec:expconfigs_specificvsgeneric}
The proposed presence masking strategies in \cref{sec:partiallylabel} were first evaluated using the \gls*{msd} dataset, with 80:20 split into train $D_{train}$ and test $D_{test}$ sets. 11 combinations of loss function and presence array/mask were investigated. For each combination, 3 specific-classifiers were trained using the completely labelled liver, pancreas, and spleen datasets respectively, and 1 generic-classifier was trained using the merged partially labelled dataset. Since the presence masking strategies can potentially affect the specific-classifiers' performance (see \cref{sec:presence_masking_complete_data}), all 11 combinations were applied to both the specific- and generic-classifiers to ensure a fair comparison. The entire setup was then repeated in 3 orthogonal slicing directions (i.e. axial, coronal, and sagittal) to identify the best performing direction for next experiment (\cref{sec:expconfigs_smalldataset}).

\subsubsection{Training with small datasets}
\label{sec:expconfigs_smalldataset}
Performance of the presence masking strategies were then evaluated with small training dataset to investigate the impact of training size on the specific- and generic-classifier approaches. From the original 80:20 train-test split, the training dataset was gradually halved to form smaller datasets 40\%, 20\%, 10\%, 5\%, and 2.5\% of the original datasets. The specific- and generic-classifiers were then trained with each of the reduced training datasets. Loss functions (5 choices for both specific- and generic-classifier) as well as slicing direction (i.e. axial) were all chosen based on results from \cref{sec:expconfigs_specificvsgeneric}.

\subsubsection{Application on human thigh \& calf datasets}
\label{sec:expconfigs_musc}
Finally, the presence masking strategies were further evaluated using the human thigh and calf datasets. To compliment the evaluation against the \gls*{msd} dataset (i.e. \cref{sec:expconfigs_specificvsgeneric}), all 11 loss function and presence array/mask combinations were again investigated with 80:20 train-test split. To ensure a fairer evaluation, an additional condition was imposed such that if multiple images exist for one subject, these images must all belong to either the training or the test set, but not both. For each combination, 2 specific-classifiers were trained using the calf-tissue and calf-muscle datasets respectively, while 1 generic-classifier was trained using the combined partially labelled calf dataset. On the other hand, only 1 specific-classifier was trained using the thigh dataset since it is completely labelled. Since \gls*{imaf} was not labelled within the entire calf dataset, a workaround was implemented: within the calf-tissue dataset, the entire region medial from \gls*{scf} was labelled as \gls*{imaf} (see \cref{fig:screenshot_and_presence_masks_calf}), but the element corresponding to \gls*{imaf} was marked as False in the presence array. This allowed the classifier to learn labeling undefined region within vicinity of the muscle groups as \gls*{imaf} instead of background.

\section{Results and discussions}
\label{sec:results_and_discussions}

\subsection{Evaluation 1: specific- versus generic-classifiers}
\label{sec:result_specificvsgeneric}

%\afterpage{
	\begin{table*}[!htbp]%
	    \caption{Dice performance of the specific- and generic-classifiers on the test set $D_{test}$, tabulated against loss function used in training the classifiers. Dice scores in bold correspond to the best performing classifier in each label class. Only classifiers trained with axial slices are shown, which is discovered to be the best performing slicing direction. Dice performance on tumor classes are also omitted. See \cref{tab:s_vs_g_result_supplement} in the supplementary material for full result.}%
		\label{tab:s_vs_g_result}
		\centering%
		\resizebox{0.95\linewidth}{!}{
		    \begin{tabular}{llllllllll}
\toprule
      & label\_class & \multicolumn{2}{l}{average} & \multicolumn{2}{l}{liver} & \multicolumn{2}{l}{pancreas} & \multicolumn{2}{l}{spleen} \\
      & classifier\_type &                   generic &         specific &          generic &         specific &          generic &         specific &          generic &         specific \\
plane & loss\_function &                           &                  &                  &                  &                  &                  &                  &                  \\
\midrule
\multirow{12}{*}{axial} & $dice_{soft}$ &                    0.0014 &           0.0000 &           0.0000 &           0.0000 &           0.0043 &           0.0000 &           0.0000 &           0.0000 \\
      & $dice_{soft}$(w/o empty) &                    0.4313 &           0.7283 &           0.0000 &           0.8606 &           0.5254 &           0.5153 &           0.7684 &           0.8089 \\
      & $dice_{log}$ &                    0.1959 &           0.5850 &           0.2278 &           0.8244 &           0.2672 &           0.1606 &           0.0925 &           0.7702 \\
      & $xent_{base}$ &                    0.0814 &  \textbf{0.8319} &           0.1717 &           0.9064 &           0.0475 &  \textbf{0.6919} &           0.0251 &           0.8973 \\
      & $xent_{base}+$$0.1*dice_{soft}$ &                    0.6886 &           0.8135 &           0.6047 &           0.8857 &           0.6408 &           0.6641 &           0.8202 &           0.8907 \\
      & $xent_{base}+$$dice_{log}$ &                    0.7637 &           0.7952 &           0.8423 &           0.8915 &           0.5933 &           0.5963 &           0.8556 &           0.8977 \\
      & $xent_{or}$ &                    0.8185 &           0.7792 &           0.9049 &           0.9045 &           0.6845 &           0.6209 &           0.8662 &           0.8121 \\
      & $xent_{or}+$$0.1*dice_{soft}$ &                    0.8148 &           0.8299 &           0.8980 &           0.9023 &           0.6860 &           0.6869 &           0.8605 &  \textbf{0.9005} \\
      & $xent_{or}+$$dice_{log}$ &                    0.8033 &           0.7799 &           0.9018 &           0.8976 &           0.6419 &           0.6222 &           0.8662 &           0.8198 \\
      & $xent_{plus}$ &                    0.8202 &           0.8107 &  \textbf{0.9097} &  \textbf{0.9080} &           0.6619 &           0.6735 &           0.8890 &           0.8506 \\
      & $xent_{plus}+$$0.1*dice_{soft}$ &           \textbf{0.8313} &           0.8210 &           0.8944 &           0.9063 &  \textbf{0.6985} &           0.6716 &  \textbf{0.9009} &           0.8851 \\
      & $xent_{plus}+$$dice_{log}$ &                    0.7979 &           0.8125 &           0.8989 &           0.9060 &           0.6528 &           0.6493 &           0.8421 &           0.8821 \\
\bottomrule
\end{tabular}

		}
		%\bigskip
		%\bigskip
		%\includegraphics[width=\linewidth,page=4]{MSD2018_MICCAI2019/seaborn_plot}
	\end{table*}
	\begin{figure*}[!ht]
		\centering
% 		\begin{subfigure}{0.99\textwidth}%{0.435\textwidth}
% 			\centering
% 			\includegraphics[width=\linewidth,page=6]{MSD2018_MICCAI2019/seaborn_plot}%page5
% 			\caption{}%{Visualization of \cref{tab:s_vs_g_result}.}
% 			\label{fig:s_vs_g_result}
% 		\end{subfigure}
% 		\begin{subfigure}{\textwidth}%{0.555\textwidth}
% 			\centering
			\includegraphics[width=\linewidth,page=1]{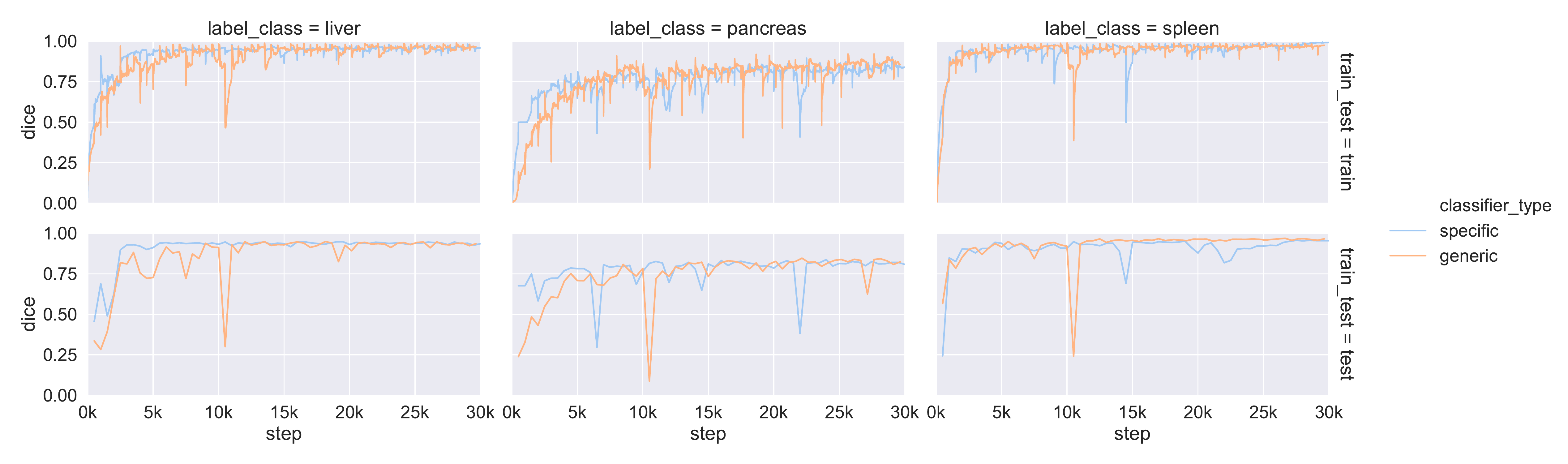}%page2
% 			\caption{}%{Training curve: specific- versus generic-classifiers}
			% \label{fig:s_vs_g_trainingcurve}
% 		\end{subfigure}
		\caption{Training curves of the specific- and generic-classifier trained with loss function $xent_{or} + 0.1 * dice_{soft}$. Train and test dice scores are plotted against training step elapsed for each label class.}
		\label{fig:s_vs_g_trainingcurve}
	\end{figure*}
%	\begin{figure}[!ht]
%		\centering
%		\includegraphics[width=\linewidth,page=5]{MSD2018_MICCAI2019/seaborn_plot}
%		\caption{Visualization of \cref{tab:s_vs_g_result}.}
%		\label{fig:s_vs_g_result}
%	\end{figure}
%	\begin{figure}[!ht]
%		\centering
%		\includegraphics[width=\linewidth,page=1]{MSD2018_trainspeed/seaborn_plot}
%		\caption{Training curve: specific- versus generic-classifiers}
%		\label{fig:s_vs_g_trainingcurve}
%	\end{figure}
%}

Results shown in \cref{tab:s_vs_g_result} indicate that soft dice loss failed altogether for both the specific- and generic-classifier. This was mitigated among the specific-classifiers when empty slices (i.e. slices without any foreground label) were removed from $D_{train}$, as seen from the row with loss function $dice_{soft}\text{(w/o empty)}$ in \cref{tab:s_vs_g_result}. However, without exposure to empty slices the classifiers then suffered from high false-positive errors, resulting in poorer precision and dice performance on $D_{test}$. This was resolved by using combinations of crossentropy and soft dice as the loss function (see \cref{eq:xentdiceloss} in the supplementary material).

As expected following arguments under \cref{sec:partiallylabel}, \cref{tab:s_vs_g_result} shows that among generic-classifiers with a crossentropy component in the loss function, inclusion of the presence mask generated in $mode_{or}$ and $mode_{plus}$ resulted in improved performance as compared to $mode_{base}$ (see \cref{fig:s_vs_g_result} in the supplementary material for visualization). This was not observed among the specific-classifiers, which indicates that the benefit of applying the presence masking strategies in training the specific-classifier (see \cref{sec:presence_masking_complete_data}) is not obvious. In another words, special handling of the false-positive errors with $mode_{or}$ and $mode_{plus}$ did not significantly improve classifiers trained with completely labelled dataset.

It worth noting that both specific- and generic-classifier failed to segment tumor classes properly (see \cref{tab:s_vs_g_result_supplement} in the supplementary material). In fact, their performance in segmenting pancreas was also mediocre at best. This is not unexpected, since these label classes are classically challenging, and tend to require handlings more sophisticated than the generic crossentropy and soft dice loss \cite{abraham_focaltversky,oktay_attentionunet}. Since this work only aims to compare performance between the specific- and generic-classifier under the same setting, improving performance of difficult label classes is omitted.

\subsubsection{Performance}
\label{sec:result_specificvsgeneric_performance}
Predicted label masks can be generated in a few ways during inference time. Where maximal accuracy is desired, this can be done using multiple best performing generic-classifiers. From the bold entries under each column of \cref{tab:s_vs_g_result}, this corresponds to the $xent_{plus}$ (best for liver) and $xent_{plus}+0.1*dice_{soft}$ (best for pancreas/spleen) generic-classifier. Multiple label maps are then predicted using these generic-classifiers, followed by masking out label classes where the corresponding generic-classifier does not perform best. Finally, a label map can be obtained by merging these masked label maps. Note that this is very similar to the conventional specific-classifiers approach, except for the extra masking step since label maps predicted by the generic-classifiers are completely labelled. Alternatively, a more efficient approach is described under \cref{sec:result_specificvsgeneric_efficiency}.

\cref{tab:s_vs_g_result,} indicates that dice score achieved by the best performing generic-classifier marginally outperformed the best specific-classifiers. This suggests that generating label maps with generic-classifier approach potentially yields better performance. However, depending on applications, this marginal performance gain may not justify the efficiency of inferencing with one single generic-classifier instead (see \cref{sec:result_specificvsgeneric_efficiency}).

\subsubsection{Efficiency}
\label{sec:result_specificvsgeneric_efficiency}
At inference time, it is motivating to predict the entire label masks with one single generic-classifier, rather than merging multiple predictions from the best performing generic-classifiers as described under \cref{sec:result_specificvsgeneric_performance}. Ideally, the selected generic-classifier should perform reasonably well across each label class, so as to minimize sacrifice on performance while attempting to improve efficiency. This can be achieved with various selection criterion, such as by choosing the generic-classifier which yields the highest average dice score. From \cref{tab:s_vs_g_result}, this corresponds to the $xent_{plus}+0.1*dice_{soft}$ generic-classifier (0.8313 on average), which performed slightly worse (0.8944) than the best performing generic-classifier (0.9097) in liver, but could nonetheless promote a good trade-off between performance per-class and the number of inferences required per-image. This could be particularly useful when computational budget during inference is limited.

On the other hand, in terms of training speed, \cref{fig:s_vs_g_trainingcurve} indicates that convergence of the generic-classifier's training and test dices closely followed the specific-classifier's. This lag was more prevailing for label class (i.e. pancreas) with distinct appearance from other classes (i.e. liver and spleen). Nonetheless, training a single generic-classifier is still arguably more efficient without exceeding the combined duration needed for training 3 specific-classifiers to convergence. In other words, training $\mathcal{O}(1)$ generic-classifiers to segment $n$ label classes is asymptotically less expensive than training $\mathcal{O}(n)$ specific-classifiers.

\subsection{Evaluation 2: training with small datasets}
\label{sec:result_smalldataset}

%\afterpage{
	\begin{figure*}[htbp]%
		\centering
		\begin{subfigure}[h]{0.95\textwidth}
			\includegraphics[width=\linewidth,page=1]{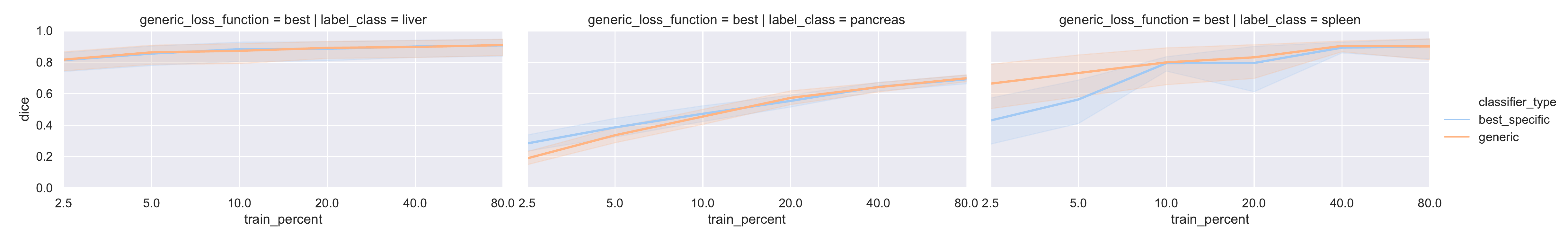}
			\caption{}
			\label{fig:smalldataset_result_bestgenericclassifier}
		\end{subfigure}
		\begin{subfigure}[h]{0.95\textwidth}
			\includegraphics[width=\linewidth,page=2]{MSD2018_smalldataset/seaborn_plot}
			\caption{}
			\label{fig:smalldataset_result_bestonavggenericclassifier}
		\end{subfigure}
		\caption{Visualization of results from \cref{sec:result_smalldataset}. See \cref{tab:smalldataset_result} in the supplementary material for full result. (\protect\subref{fig:smalldataset_result_bestgenericclassifier}) Dice score achieved by the best performing specific- and generic-classifier are plotted against size of $D_{train}$ (i.e. bold entries in \cref{tab:smalldataset_result}); (\protect\subref{fig:smalldataset_result_bestonavggenericclassifier}) same as \protect\subref{fig:smalldataset_result_bestgenericclassifier} but with the best-on-average generic-classifiers instead. These are generic-classifiers scoring the highest dice score on-average for each $D_{train}$ size, i.e. from \cref{tab:smalldataset_result}, $xent_{or}+dice_{log}$ trained generic-classifiers at 2.5\%, 5\%, and 10\% $D_{train}$, $xent_{plus}$ at 20\%, and $xent_{plus}+0.1*dice_{soft}$ at 40\% and 80\%.}
		\label{fig:smalldataset_result}
	\end{figure*}
%}

As a sanity check, \cref{fig:smalldataset_result} shows that performance of the classifiers improved with larger $D_{train}$. This observation holds true for both specific- and generic-classifier, which is as expected.

\subsubsection{Performance}
\label{sec:result_smalldataset_performance}
\cref{fig:smalldataset_result_bestgenericclassifier} shows that among label classes with similar appearance (i.e. liver and spleen), performance of the best generic-classifier often outperformed the best specific-classifier. This was increasingly apparent as size of $D_{train}$ shrinks, particularly for label class from the smaller dataset (i.e. spleen). Label classes with distinct appearance (i.e. pancreas), however, did not benefit from this performance boost.

This observed behavior is very likely linked to transfer learning, where classifiers are pre-trained using large dataset prior to fine-tuning for target application with smaller dataset. While in this case it is likely that the spleen specific-classifier's performance can be boosted by simply pre-training the classifier with liver dataset, it is not so straightforward when, say, a kidney dataset is simultaneously available and can potentially improve the spleen specific-classifier's performance too. The generic-classifier approach extends transfer learning's capability such that the pre-training step leverages on multiple dataset simultaneously, i.e. without manual grouping of label classes based on their appearance. 

\subsubsection{Efficiency}
\label{sec:result_smalldataset_efficiency}
With smaller $D_{train}$ sizes, the strategy described in \cref{sec:result_specificvsgeneric_efficiency} to improve inference efficiency failed, as shown in \cref{fig:smalldataset_result_bestonavggenericclassifier}. Dice score achieved by the selected, best-on-average generic-classifiers at each $D_{train}$ size were often low as compared to the best generic-classifiers. This could potentially be resolved with minor amendment to the aforementioned strategy. For example, the generic-classifier performing best on-average in segmenting liver and spleen could be used in conjunction with the best pancreas specific-classifier during inference. Such strategy could still improve overall efficiency of the inference process, but at the expense of additional workflow complexity.

Meanwhile, as shown in the supplemented \cref{tab:smalldataset_result}, the average dice score between specific- and generic-classifier matched closely for each $D_{train}$ size. Generic-classifiers could potentially be trained with class weightages additionally assigned to each label class in the loss function, such that the generic-classifier's dice performance is well-balanced across each label class, while maintaining the same average dice score. Since the optimal set of class weightages may require exhaustive tuning, this exploration is omitted here.

\subsection{Evaluation 3: application on human thigh \& calf datasets}
\label{sec:result_musc}

%\afterpage{
%	\begin{table}[htbp]%
%		\caption{Results: calf and thigh}%
%		\centering%
%		\resizebox{\linewidth}{!}{\input{figures/MUSC/latex_table_average}}
%		\label{tab:musc_average}
%	\end{table}
%	\begin{table*}[htbp]%
%		\caption{Results: calf}%
%		\centering%
%		\resizebox{\linewidth}{!}{\input{figures/MUSC/latex_table_calf}}
%		\label{tab:musc_calf}
%	\end{table*}
%	\begin{table*}[htbp]%
%		\caption{Results: thigh}%
%		\centering%
%		\resizebox{\linewidth}{!}{\input{figures/MUSC/latex_table_thigh}}
%		\label{tab:musc_thigh}
%	\end{table*}
	\begin{figure*}[htbp]
		\centering
% 		\begin{subfigure}[t]{\linewidth}
% 			\centering
% 			\includegraphics[width=\linewidth,page=3]{MUSC/seaborn_plot}
% 			\caption{}
% 			\label{fig:musc_result_calf_best}
% 		\end{subfigure}
% 		\hfill
% 		\begin{subfigure}[t]{\linewidth}
% 			\centering
			\includegraphics[width=\linewidth,page=4]{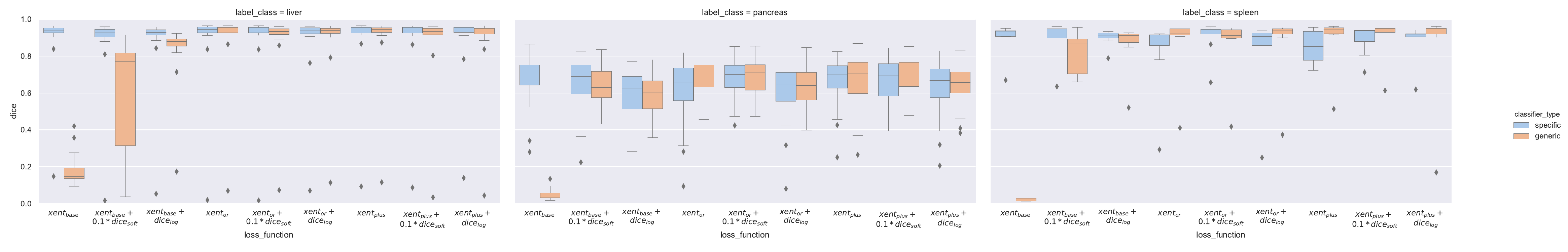}
% 			\caption{}
			% \label{fig:musc_result_calf_bestonavg}
% 		\end{subfigure}
		%\hfill
		%\begin{subfigure}[t]{\linewidth}
		%	\centering
		%	\includegraphics[width=\linewidth,page=5]{MUSC/seaborn_plot}
		%	\caption{}
		%	\label{fig:musc_result_thigh_best}
		%\end{subfigure}
		\caption{Visualization of results from \cref{sec:result_musc}, comparing Dice scores achieved by the best specific-classifiers and best-on-average generic-classifiers in segmenting the calf. \textit{pl+edl} represents the pixel-wise union of \textit{peroneus longus} and \textit{extensor digitorum longus}. See \cref{tab:musc_result} in the supplementary material for full result.}%; \protect\subref{fig:musc_result_thigh_best}: best specific-classifiers, thigh}
% 		\label{fig:musc_result}
        \label{fig:musc_result_calf_bestonavg}
	\end{figure*}

In general, evaluation results were in good agreement with \cref{sec:result_specificvsgeneric} for the \gls*{msd} dataset. From \cref{fig:musc_result_calf_bestonavg}, the best-on-average generic-classifier (trained with $xent_{plus}+dice_{log}$) was capable of delivering performance on par with that of the best specific-classifiers, suggesting that efficiency can be improved by inferencing all label classes with the selected generic-classifier only. For certain label classes, the best generic-classifiers performed marginally better than the best-on-average generic-classifier (see the supplemented \cref{tab:musc_result}). However, these generic-classifiers were not better than the best specific-classifiers in general, which is expected since label classes annotated in the calf-tissue and calf-muscle datasets mostly do not share similar appearance (see \cref{fig:screenshot_and_presence_masks_calf}).

It is again observed that $mode_{or}$ and $mode_{plus}$ performed better in general than $mode_{base}$ among the calf generic-classifiers, while such advantage was not observed among the calf and thigh specific-classifiers (see \cref{tab:musc_result}). Screenshot of label maps predicted by the calf generic-classifiers and thigh specific-classifiers demonstrate this visually (see \cref{fig:musc_screenshot} in the supplementary material). Among the calf generic-classifiers, loss functions with $xent_{or}$ or $xent_{plus}$ components were in general better than $xent_{base}$ in isolating the muscle groups from the surrounding \gls*{imaf}, which was undefined among all ground-truth label map within the calf dataset. Such advantage was not observed among the thigh specific-classifiers.

In terms of absolute performance of the classifiers, dice score was acceptable for most classes given the small size of $D_{train}$. The only exception was the vessels, which is again a small \gls*{roi} that potentially requires special handling just like tumor classes from the \gls*{msd} dataset. Additionally, both calf generic- and specific-classifiers were unable to delineate between the 2 muscle groups \textit{peroneus longus} and \textit{extensor digitorum longus} well, but were capable of identifying union of the two label classes, which is represented by $pl+edl$ on \cref{fig:musc_result_calf_bestonavg,tab:musc_result}.

\subsection{Potential improvements}
\label{sec:potential_improvements}

\subsubsection{Compatibility with existing \gls*{dl} methodologies}
This work mainly aims at comparing the performance between existing methods (i.e. specific-classifier) and the proposed generic-classifier approaches, rather than pushing performance of the generic-classifier itself to achieve state-of-the-art semantic segmentation performance. However, it is possible to incorporate the proposed presence masking strategies into many other existing \gls*{dl} methodologies for better performance. This includes adopting the DeepMedic \cite{kamnitsas_deepmedic} and HyperDense-Net \cite{dolz2019_hyperdense} architectures which excelled in brain lesion segmentation, or the Attention U-Net \cite{oktay_attentionunet} originally proposed for pancreas segmentation. It is also common to see fully-connected \gls*{crf} being appended to the network output to improve delineation near \gls*{roi} boundaries \cite{papandreou2015_weaksemisupervised,kamnitsas_deepmedic}. Additionally, other strategies such as various data augmentation schemes \cite{zhao2019_augmentation} or 2D-to-3D segmentation map by majority voting \cite{zhou2016_majorityvoting} should also be directly applicable.

Note that while the proposed feature array/masks were only validated using dice and crossentropy loss, incorporation to other loss function is trivial in some cases. For instance, the \gls*{ftl} \cite{abraham_focaltversky} is defined as followed:
\begin{equation}
J_{FTL}(\boldsymbol{\theta}) = \sum_{c}^{C}\nolimits (1-{TI}^{c})^{1/\gamma}
\label{eq:ftl}
\end{equation}
with ${TI}_{c}$ being the Tversky index \cite{tversky1977_tverskyindex} between the ground-truth and predicted label map for label class $c$. This can be easily modified to incorporate the presence array $\textbf{k}=\{k^{c}\}$:
\begin{equation}
J_{FTL}(\boldsymbol{\theta}) = \frac{\sum_{c}^{C}\nolimits k^{c}(1-{TI}^{c})^{1/\gamma}}{\sum_{c}^{C}\nolimits k^{c}}
\label{eq:masked_ftl}
\end{equation}
Meanwhile, other loss functions with pixel aggregation terms, such as the \gls*{mae} or L1 loss which is proposed as a noise-robust alternative to crossentropy loss \cite{ghosh2017_robustmae}:
\begin{align}
	J_{MAE}(\boldsymbol{\theta}) &= - \sum_{m}^{M}\nolimits \sum_{c}^{C}\nolimits \left| y_{m}^c - p_{m}^c \right| \nonumber \\
	&= - \sum_{m}^{M}\nolimits 2 - 2 p_{m}^{c^\prime},\quad c^\prime = y_{m}
	\label{eq:maeloss}
\end{align}
can be handled in a fashion similar to \cref{eq:masked_xentloss}:
\begin{equation}
	J_{MAE}(\boldsymbol{\theta}) = - \sum_{m}^{M}\nolimits w_{m} \cdot (2 - 2 p_{m}^{c^\prime})
	\label{eq:masked_maeloss}
\end{equation}

\subsubsection{Semi-supervised learning}
%As an overview, semi-supervised learning aims to improve classifier's performance by training on both labelled and unlabelled data \cite{zhu_introtosemisupervised}. For semantic segmentation, weak labels (e.g. point, bounding-box, or image level labels \cite{papandreou2015_weaksemisupervised,bearman_whatsthepoint,rajchl2016_bboxsegmentation}) may also be leveraged besides the strong, pixel-level label maps. Weakly-labelled and unlabelled images typically require additional processing. For example, in self-supervised learning with the \gls*{em} algorithm \cite{dempster_em}, the training process alternates between predicting strong labels for weakly- and un-labelled feature images, and optimizing the classifier using both manually-annotated and estimated strong labels as ground-truth.

Technically, the proposed presence masking strategies are much closer to supervised learning than it is to semi-supervised learning. All labels involved in the training are strongly labelled, and classifiers do not learn from the inferred labels. The only exception is perhaps when the classifier makes false positive error at unlabelled pixels, which contribute to the loss function (with presence masks under $mode_{or}$ and $mode_{plus}$) even without being annotated with any ground-truth label class. Nevertheless, it is possible to extend the proposed method to support semi-supervised learning scenarios. Since predicted label maps for the partially labelled training images are completely labelled, they can be merged into $D_{train}$. Classifier can then be iteratively trained until certain termination conditions are met, similar to the \gls*{em} approach.

% \section{Conclusion}
% \label{sec:conclusion}
% In this paper, we proposed presence masking strategies to enable training of classifiers using multiple datasets of partially labelled images, while remaining in a fully-supervised learning regime. These strategies apply a weightage mask to the loss function, and differ mainly in how false positive/negative errors are weighted. We evaluated these strategies using the \gls*{msd} dataset. By reducing the number of classifiers required, we successfully demonstrated that both training and inference efficiency can be improved from $\mathcal{O}(n)$ to $\mathcal{O}(1)$, where $n$ is the number of label classes. We also showed that training a classifier with a combination of multiple small, partially labelled datasets improved performance, as compared to training multiple classifiers using each dataset separately. The proposed strategies were then applied to segmentation of MRI images of human lower limb, from which we observed that the result was in good agreement with those on segmenting the abdominal organs from the \gls*{msd} dataset. Future work will focus on evaluating the strategies with other \gls*{dl} methodologies, which include but are not limited to other network architectures and similarly-structured loss functions.

% \appendices

\section*{Acknowledgment}
The computational work for this article was partially performed on resources of the National Supercomputing Centre, Singapore (https://www.nscc.sg). The authors also gratefully acknowledge the support of NVIDIA Corporation with the donation of the Titan V GPU used for inferences in this research. All codes are publicly available at https://github.com/wong-ck/DeepSegment.

\bibliography{refs.bib}{}
\bibliographystyle{IEEEtran}

\clearpage

% supplementary materials
\setcounter{page}{1}
\setcounter{section}{0}
\setcounter{table}{0}
\setcounter{figure}{0}
\setcounter{equation}{0}

\renewcommand{\thepage}{S\arabic{page}}
\renewcommand{\thesection}{S\arabic{section}}
\renewcommand{\thetable}{S\arabic{table}}
\renewcommand{\thefigure}{S\arabic{figure}}
\renewcommand{\theequation}{S\arabic{equation}}

\section{Supplementary materials}
\subsection{Overview of supervised semantic segmentation}
\label{sec:method_overview}
From an input feature image $\mathbf{x}=\{x_{m}\}$, a \gls*{cnn} classifier $f(\mathbf{x},\boldsymbol{\theta})$ estimates the per-class probability $\mathbf{p}=\{p_m^{c}\}$ that $x_{m}$ should be assigned the ground-truth label class c:
\begin{equation}
p_{m}^{c} = \text{P}(y_{m}=c\mid\mathbf{x},\boldsymbol{\theta})
\label{eq:classprobabilitymap}
\end{equation}
from which the predicted label map $\mathbf{\hat{y}}=\{\hat{y}_m\}$ is generated:
\begin{equation}
\hat{y}_m=\argmax_{c}\nolimits p_{m}^{c}
\label{eq:predictedmap}
\end{equation}

\gls*{cnn} classifier training may take multiple forms, which include supervised learning when there exists a training dataset $D_{train}=\{(\mathbf{x},\mathbf{y})_{n}\}, n\in\{1,\dots,N\}$ with $N$ annotated image pairs $(\mathbf{x},\mathbf{y})$. This involves optimizing the model parameters $\boldsymbol{\theta}$ to minimize the value of a selected loss function $J(\boldsymbol{\theta})$, often by back-propagating the gradient $\frac{dJ}{d\boldsymbol{\theta}}$ with some variant of the gradient decent algorithm.

Loss function essentially measures difference between the ground-truth label \& current predictions, and is often chosen based on the problem in hand. In solving classification problems, it is common to start with crossentropy loss:
\begin{equation}
J_{xent}(\boldsymbol{\theta}) = - \sum_{m}^{M}\nolimits \sum_{c}^{C}\nolimits y_{m}^c \log p_{m}^c
\label{eq:xentloss}
\end{equation}

Meanwhile, performance of \gls*{cnn} classifier in semantic segmentation is often evaluated using some form of \gls*{iou} matrix. It is hence intuitive to select \gls*{iou}-derived losses for semantic segmentation problems. One popular option is the soft dice loss \cite{milletari2016_vnet}:
\begin{equation}
J_{dice_{soft}}(\boldsymbol{\theta}) = \sum_{c}^{C}\nolimits (1 - softdice^{c})
\label{eq:softdiceloss}
\end{equation}
which is derived from the soft dice score for every class $c$:
\begin{equation}
softdice^{c} = \frac{ 2 \cdot \sum_{m}^{M} (y_{m}^{c} \cdot p_{m}^{c}) }{ \sum_{m}^{M} y_{m}^{c} + \sum_{m}^{M} p_{m}^{c} }
\label{eq:dice}
\end{equation}

%For example, with the soft dice score defined for every class $c$ as:
%\begin{equation}
%softdice^{c} = \frac{ 2 \cdot \sum_{m}^{M} (y_{m}^{c} \cdot p_{m}^{c}) }{ \sum_{m}^{M} y_{m}^{c} + \sum_{m}^{M} p_{m}^{c} }
%\label{eq:dice}
%\end{equation}
%the soft dice loss \cite{milletari2016_vnet} can be derived:
%\begin{equation}
%J_{dice_{soft}}(\boldsymbol{\theta}) = \sum_{c}^{C}\nolimits (1 - softdice^{c})
%\label{eq:softdiceloss}
%\end{equation}
%which is a popular choice in semantic segmentation.

Minimizing the soft dice loss is effectively equivalent to maximizing the \gls*{iou}-based soft dice score. Intuitively, other loss functions derived in a similar manner should also achieve the same effect, such as minimizing the negative log of the soft dice score:
\begin{equation}
J_{dice_{log}}(\boldsymbol{\theta}) = - \sum_{c}^{C}\nolimits \log{softdice^{c}}
\label{eq:logdiceloss}
\end{equation}

With both crossentropy and soft dice loss coming with their own strengths and weaknesses, some studies adopt a merged version of the two losses \cite{isensee2018_nnunet,khened2019_dualloss}:
\begin{equation}
J_{xentdice}(\boldsymbol{\theta}) = a \cdot J_{xent}(\boldsymbol{\theta}) + b \cdot J_{dice}(\boldsymbol{\theta})
\label{eq:xentdiceloss}
\end{equation}
with $a$ and $b$ controlling relative contribution of the two loss components.

\newpage

\subsection{Dataset details and full evaluation results}
\begin{table}[!ht]%
	\centering%
	\caption{Detail of human thigh and calf datasets}%
	\label{tab:musc_dataset_desc}
	\resizebox{\linewidth}{!}{\begin{tabular}{lllrl}
\toprule
     dataset &      dimensions &    resolutions (mm) &  count & image content \\
\midrule
 calf-muscle &  (384, 308, 44) &  (0.47, 0.47, 3.00) &     18 &   single calf \\
 calf-muscle &  (384, 308, 88) &  (0.47, 0.47, 1.00) &      4 &   single calf \\
 calf-tissue &  (704, 496, 25) &  (0.54, 0.54, 7.00) &     42 &   both calves \\
       thigh &  (512, 512, 32) &  (0.39, 0.39, 5.00) &     40 &  single thigh \\
\bottomrule
\end{tabular}
}
\end{table}

% We include as supplementary material the full results of our experimental evaluations under \cref{sec:results_and_discussions}.
\begin{table*}[!htbp]
    \caption{Full result of \cref{sec:result_specificvsgeneric}. This is an extension of \cref{tab:s_vs_g_result} in the main text, with dice scores for the tumor label classes, as well as classifiers trained using images sliced in the coronal and sagittal directions.}%
	\label{tab:s_vs_g_result_supplement}
	\centering%
	\resizebox{\linewidth}{!}{\begin{tabular}{llllllllllllll}
\toprule
         & label\_class & \multicolumn{2}{l}{average} & \multicolumn{2}{l}{liver} & \multicolumn{2}{l}{pancreas} & \multicolumn{2}{l}{spleen} & \multicolumn{2}{l}{liver\_tumor} & \multicolumn{2}{l}{pancreas\_tumor} \\
         & classifier\_type &               generic &         specific &          generic &         specific &          generic &         specific &          generic &         specific &          generic &         specific &          generic &         specific \\
plane & loss\_function &                       &                  &                  &                  &                  &                  &                  &                  &                  &                  &                  &                  \\
\midrule
\multirow{12}{*}{axial} & $dice_{soft}$ &                0.0316 &           0.0308 &           0.0000 &           0.0000 &           0.0043 &           0.0000 &           0.0000 &           0.0000 &           0.1538 &           0.1538 &           0.0000 &           0.0000 \\
         & $dice_{soft}$(w/o empty) &                0.2595 &           0.4677 &           0.0000 &           0.8606 &           0.5254 &           0.5153 &           0.7684 &           0.8089 &           0.0037 &           0.1538 &           0.0000 &           0.0000 \\
         & $dice_{log}$ &                0.1485 &           0.3571 &           0.2278 &           0.8244 &           0.2672 &           0.1606 &           0.0925 &           0.7702 &           0.1538 &           0.0300 &           0.0009 &           0.0004 \\
         & $xent_{base}$ &                0.0642 &  \textbf{0.6198} &           0.1717 &           0.9064 &           0.0475 &  \textbf{0.6919} &           0.0251 &           0.8973 &           0.0642 &           0.4849 &           0.0126 &           0.1185 \\
         & $xent_{base}+$$0.1*dice_{soft}$ &                0.4885 &           0.5227 &           0.6047 &           0.8857 &           0.6408 &           0.6641 &           0.8202 &           0.8907 &           0.3767 &           0.1732 &           0.0002 &           0.0000 \\
         & $xent_{base}+$$dice_{log}$ &                0.5673 &           0.6040 &           0.8423 &           0.8915 &           0.5933 &           0.5963 &           0.8556 &           0.8977 &           0.4070 &  \textbf{0.4933} &           0.1385 &           0.1411 \\
         & $xent_{or}$ &                0.6116 &           0.5777 &           0.9049 &           0.9045 &           0.6845 &           0.6209 &           0.8662 &           0.8121 &           0.4475 &           0.4341 &           0.1551 &           0.1169 \\
         & $xent_{or}+$$0.1*dice_{soft}$ &                0.6024 &           0.6165 &           0.8980 &           0.9023 &           0.6860 &           0.6869 &           0.8605 &  \textbf{0.9005} &           0.4044 &           0.4851 &           0.1629 &           0.1078 \\
         & $xent_{or}+$$dice_{log}$ &                0.6071 &           0.5895 &           0.9018 &           0.8976 &           0.6419 &           0.6222 &           0.8662 &           0.8198 &  \textbf{0.4718} &           0.4791 &           0.1539 &           0.1289 \\
         & $xent_{plus}$ &       \textbf{0.6191} &           0.6073 &  \textbf{0.9097} &  \textbf{0.9080} &           0.6619 &           0.6735 &           0.8890 &           0.8506 &           0.4437 &           0.4450 &           0.1911 &  \textbf{0.1592} \\
         & $xent_{plus}+$$0.1*dice_{soft}$ &                0.6105 &           0.5780 &           0.8944 &           0.9063 &  \textbf{0.6985} &           0.6716 &  \textbf{0.9009} &           0.8851 &           0.4097 &           0.4164 &           0.1490 &           0.0106 \\
         & $xent_{plus}+$$dice_{log}$ &                0.6105 &           0.6076 &           0.8989 &           0.9060 &           0.6528 &           0.6493 &           0.8421 &           0.8821 &           0.4521 &           0.4742 &  \textbf{0.2065} &           0.1265 \\
\cline{1-14}
\multirow{12}{*}{coronal} & $dice_{soft}$ &                0.1531 &           0.0308 &           0.0000 &           0.0000 &           0.0000 &           0.0000 &           0.7618 &           0.0000 &           0.0039 &           0.1538 &           0.0000 &           0.0000 \\
         & $dice_{soft}$(w/o empty) &                0.3342 &           0.4512 &           0.3511 &           0.8505 &           0.4809 &           0.4687 &           0.8360 &           0.7828 &           0.0032 &           0.1538 &           0.0000 &           0.0000 \\
         & $dice_{log}$ &                0.2118 &           0.3860 &           0.3083 &           0.7083 &           0.3554 &           0.4021 &           0.2414 &           0.7387 &           0.1538 &           0.0654 &           0.0000 &           0.0156 \\
         & $xent_{base}$ &                0.0978 &           0.5273 &           0.0750 &           0.8674 &           0.0336 &           0.6144 &           0.2468 &           0.7983 &           0.1149 &           0.3181 &           0.0184 &           0.0384 \\
         & $xent_{base}+$$0.1*dice_{soft}$ &                0.3335 &           0.5499 &           0.1623 &  \textbf{0.8856} &           0.4378 &           0.5981 &           0.8352 &           0.8174 &           0.2323 &  \textbf{0.4085} &           0.0002 &           0.0399 \\
         & $xent_{base}+$$dice_{log}$ &                0.5113 &           0.4929 &           0.7909 &           0.8238 &           0.5413 &           0.4829 &           0.8349 &           0.7866 &  \textbf{0.3189} &           0.3461 &           0.0706 &           0.0255 \\
         & $xent_{or}$ &                0.5221 &           0.5150 &  \textbf{0.8734} &           0.8775 &           0.5379 &           0.5675 &           0.8311 &           0.7950 &           0.3137 &           0.2872 &           0.0542 &           0.0480 \\
         & $xent_{or}+$$0.1*dice_{soft}$ &       \textbf{0.5483} &  \textbf{0.5561} &           0.8701 &           0.8552 &           0.6263 &  \textbf{0.6175} &           0.8663 &           0.8624 &           0.3188 &           0.3843 &           0.0601 &  \textbf{0.0610} \\
         & $xent_{or}+$$dice_{log}$ &                0.5130 &           0.5217 &           0.8229 &           0.8574 &           0.5544 &           0.5550 &           0.7868 &           0.8400 &           0.2970 &           0.3049 &  \textbf{0.1039} &           0.0513 \\
         & $xent_{plus}$ &                0.5461 &           0.5141 &           0.8570 &           0.8598 &  \textbf{0.6319} &           0.5949 &           0.8516 &           0.7877 &           0.3039 &           0.2723 &           0.0861 &           0.0557 \\
         & $xent_{plus}+$$0.1*dice_{soft}$ &                0.5476 &           0.5369 &           0.8593 &           0.8744 &           0.6212 &           0.5538 &  \textbf{0.9001} &  \textbf{0.8831} &           0.3119 &           0.3640 &           0.0455 &           0.0091 \\
         & $xent_{plus}+$$dice_{log}$ &                0.4978 &           0.5183 &           0.8409 &           0.8213 &           0.5793 &           0.5603 &           0.7906 &           0.7923 &           0.2027 &           0.3718 &           0.0754 &           0.0456 \\
\cline{1-14}
\multirow{12}{*}{sagittal} & $dice_{soft}$ &                0.0240 &           0.0308 &           0.0000 &           0.0000 &           0.0047 &           0.0000 &           0.0000 &           0.0000 &           0.1154 &           0.1538 &           0.0000 &           0.0000 \\
         & $dice_{soft}$(w/o empty) &                0.3265 &           0.3904 &           0.3554 &           0.7991 &           0.5387 &           0.5254 &           0.7351 &           0.4736 &           0.0031 &           0.1538 &           0.0000 &           0.0000 \\
         & $dice_{log}$ &                0.1737 &           0.3154 &           0.2026 &           0.5913 &           0.2946 &           0.4804 &           0.2927 &           0.4285 &           0.0787 &           0.0570 &           0.0000 &           0.0199 \\
         & $xent_{base}$ &                0.0663 &  \textbf{0.5397} &           0.1111 &           0.8164 &           0.0133 &  \textbf{0.6631} &           0.0749 &           0.8138 &           0.1241 &           0.3412 &           0.0084 &           0.0639 \\
         & $xent_{base}+$$0.1*dice_{soft}$ &                0.3734 &           0.4987 &           0.1828 &  \textbf{0.8329} &           0.5870 &           0.6284 &           0.7780 &           0.6854 &           0.3187 &           0.3467 &           0.0003 &           0.0000 \\
         & $xent_{base}+$$dice_{log}$ &                0.4680 &           0.4806 &           0.7341 &           0.7357 &           0.5168 &           0.5755 &           0.7514 &           0.7664 &           0.2781 &           0.2076 &           0.0598 &  \textbf{0.1178} \\
         & $xent_{or}$ &                0.5201 &           0.5177 &           0.7815 &           0.7573 &           0.6083 &           0.5909 &           0.8127 &  \textbf{0.9061} &           0.3489 &           0.2740 &           0.0491 &           0.0601 \\
         & $xent_{or}+$$0.1*dice_{soft}$ &                0.5557 &           0.4907 &           0.8219 &           0.7640 &           0.6329 &           0.6364 &           0.8486 &           0.7010 &           0.3155 &  \textbf{0.3522} &  \textbf{0.1598} &           0.0000 \\
         & $xent_{or}+$$dice_{log}$ &                0.4932 &           0.4895 &           0.7415 &           0.7382 &           0.6162 &           0.6284 &           0.7933 &           0.7693 &           0.2548 &           0.2250 &           0.0602 &           0.0863 \\
         & $xent_{plus}$ &                0.5170 &           0.5310 &           0.7849 &           0.7842 &  \textbf{0.6417} &           0.6293 &           0.7873 &           0.8363 &           0.2854 &           0.3317 &           0.0859 &           0.0735 \\
         & $xent_{plus}+$$0.1*dice_{soft}$ &       \textbf{0.5670} &           0.5215 &  \textbf{0.8221} &           0.7623 &           0.6398 &           0.6412 &  \textbf{0.8598} &           0.8743 &  \textbf{0.3554} &           0.3270 &           0.1581 &           0.0029 \\
         & $xent_{plus}+$$dice_{log}$ &                0.4845 &           0.5146 &           0.6935 &           0.7143 &           0.5997 &           0.6521 &           0.7929 &           0.8340 &           0.2523 &           0.2867 &           0.0842 &           0.0861 \\
\bottomrule
\end{tabular}
}
\end{table*}

\begin{figure*}[!htb]
	\centering
	\includegraphics[width=\linewidth,page=6]{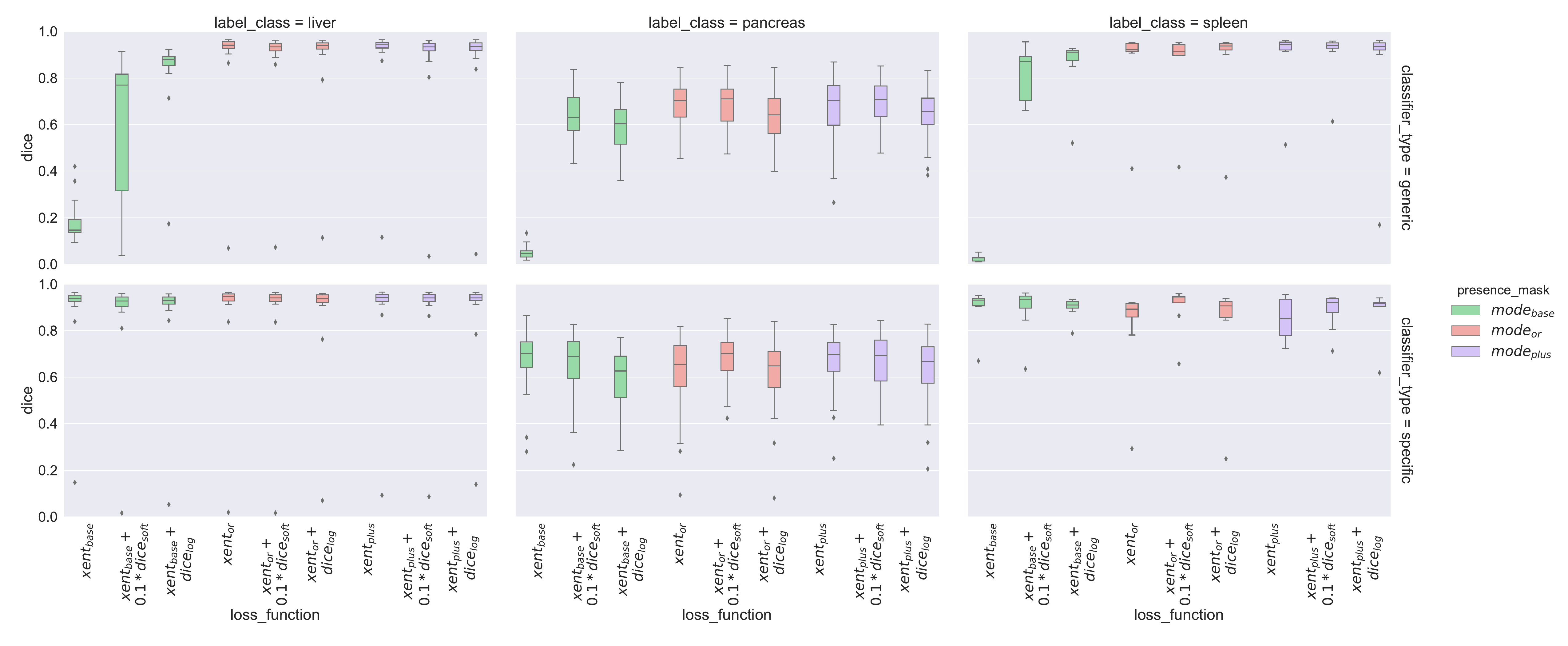}%page5
	\caption{Box and whisker plots of the result from \cref{tab:s_vs_g_result_supplement}. Dice performances of the specific-/generic-classifier pairs are plotted against the loss function they are trained with. Loss functions without any crossentropy component are omitted due to poor performance}
	\label{fig:s_vs_g_result}
\end{figure*}

%\begin{figure*}[!htbp]
%	\centering
%	\begin{subfigure}{0.49\textwidth}%{0.435\textwidth}
%		\centering
%		\includegraphics[width=\linewidth,page=7]{MSD2018_MICCAI2019/seaborn_plot}%page5
%		\caption{}%{Visualization of \cref{tab:s_vs_g_result}.}
%		\label{fig:s_vs_g_result_candidate2}
%	\end{subfigure}
%	\begin{subfigure}{0.49\textwidth}%{0.555\textwidth}
%		\centering
%		\includegraphics[width=\linewidth,page=2]{MSD2018_trainspeed/seaborn_plot}%page2
%		\caption{}%{Training curve: specific- versus generic-classifiers}
%		\label{fig:s_vs_g_trainingcurve_candidate2}
%	\end{subfigure}
%	\caption{Candidate 2 of (\protect\subref{fig:s_vs_g_result_candidate2}) \cref{fig:s_vs_g_result}, bar chart instead of box-and-whisker plot is also possible; (\protect\subref{fig:s_vs_g_trainingcurve_candidate2}) \cref{fig:s_vs_g_trainingcurve}}
%\end{figure*}

\begin{table*}[!htbp]%
	\caption{Dice performance of the specific- and generic-classifiers on $D_{test}$ as size of the training set $D_{train}$ shrinks from 80\% to 2.5\%. Dice scores in bold correspond to the best performing loss function in each combination of $D_{train}$ size, classifier type (generic or specific), and label class.}%
	\label{tab:smalldataset_result}
    \centering%
	\resizebox{\linewidth}{!}{\begin{tabular}{llllllllllllll}
\toprule
       & classifier\_type & \multicolumn{6}{l}{generic} & \multicolumn{6}{l}{specific} \\
       & train\_percent &          2.5  &          5  &          10 &          20 &          40 &          80 &          2.5  &          5  &          10 &          20 &          40 &          80 \\
label\_class & loss\_function &                  &                  &                  &                  &                  &                  &                  &                  &                  &                  &                  &                  \\
\midrule
\multirow{8}{*}{average} & $xent_{base}$ &              - &              - &              - &              - &              - &              - &           0.4921 &           0.5405 &           0.6597 &           0.7088 &           0.7282 &  \textbf{0.8319} \\
       & $xent_{base}+0.1*dice_{soft}$ &              - &              - &              - &              - &              - &              - &           0.4433 &           0.5613 &  \textbf{0.7046} &           0.6940 &           0.7686 &           0.8135 \\
       & $xent_{base}+dice_{log}$ &              - &              - &              - &              - &              - &              - &           0.3401 &           0.5040 &           0.6762 &           0.6492 &           0.7204 &           0.7952 \\
       & $xent_{plus}$ &           0.5027 &           0.6065 &           0.6770 &  \textbf{0.7614} &           0.7883 &           0.8202 &           0.4505 &           0.5648 &           0.6122 &           0.7052 &           0.7531 &           0.8107 \\
       & $xent_{or}+0.1*dice_{soft}$ &           0.4480 &           0.5972 &           0.6731 &           0.7183 &           0.8026 &           0.8148 &  \textbf{0.4933} &  \textbf{0.6010} &           0.6988 &  \textbf{0.7335} &  \textbf{0.8041} &           0.8299 \\
       & $xent_{or}+dice_{log}$ &  \textbf{0.5268} &  \textbf{0.6258} &  \textbf{0.6904} &           0.7341 &           0.7644 &           0.8033 &              - &              - &              - &              - &              - &              - \\
       & $xent_{plus}+0.1*dice_{soft}$ &           0.4512 &           0.5719 &           0.6887 &           0.7375 &  \textbf{0.8116} &  \textbf{0.8313} &              - &              - &              - &              - &              - &              - \\
       & $xent_{plus}+dice_{log}$ &           0.4928 &           0.6252 &           0.6706 &           0.7404 &           0.7852 &           0.7979 &              - &              - &              - &              - &              - &              - \\
\cline{1-14}
\multirow{8}{*}{liver} & $xent_{base}$ &              - &              - &              - &              - &              - &              - &           0.7632 &           0.8380 &           0.8568 &           0.8681 &           0.8740 &           0.9064 \\
       & $xent_{base}+0.1*dice_{soft}$ &              - &              - &              - &              - &              - &              - &           0.7391 &           0.8331 &           0.8793 &  \textbf{0.8860} &           0.8856 &           0.8857 \\
       & $xent_{base}+dice_{log}$ &              - &              - &              - &              - &              - &              - &           0.8095 &           0.7681 &           0.8579 &           0.8656 &           0.8766 &           0.8915 \\
       & $xent_{plus}$ &  \textbf{0.8170} &  \textbf{0.8633} &  \textbf{0.8729} &  \textbf{0.8918} &  \textbf{0.8974} &  \textbf{0.9097} &           0.7559 &           0.8546 &           0.8240 &           0.8830 &  \textbf{0.9003} &  \textbf{0.9080} \\
       & $xent_{or}+0.1*dice_{soft}$ &           0.6837 &           0.8528 &           0.8582 &           0.8686 &           0.8879 &           0.8980 &  \textbf{0.8136} &  \textbf{0.8553} &  \textbf{0.8835} &           0.8856 &           0.8950 &           0.9023 \\
       & $xent_{or}+dice_{log}$ &           0.7282 &           0.8171 &           0.8302 &           0.8810 &           0.8706 &           0.9018 &              - &              - &              - &              - &              - &              - \\
       & $xent_{plus}+0.1*dice_{soft}$ &           0.7502 &           0.8292 &           0.8570 &           0.8823 &           0.8886 &           0.8944 &              - &              - &              - &              - &              - &              - \\
       & $xent_{plus}+dice_{log}$ &           0.7643 &           0.8219 &           0.8621 &           0.8784 &           0.8678 &           0.8989 &              - &              - &              - &              - &              - &              - \\
\cline{1-14}
\multirow{8}{*}{pancreas} & $xent_{base}$ &              - &              - &              - &              - &              - &              - &  \textbf{0.2835} &           0.3223 &           0.4261 &  \textbf{0.5542} &  \textbf{0.6443} &  \textbf{0.6919} \\
       & $xent_{base}+0.1*dice_{soft}$ &              - &              - &              - &              - &              - &              - &           0.2412 &           0.3415 &  \textbf{0.4723} &           0.5225 &           0.5865 &           0.6641 \\
       & $xent_{base}+dice_{log}$ &              - &              - &              - &              - &              - &              - &           0.2107 &           0.2950 &           0.3766 &           0.3965 &           0.4834 &           0.5963 \\
       & $xent_{plus}$ &           0.1150 &           0.2907 &           0.3832 &           0.5608 &           0.6145 &           0.6619 &           0.2210 &           0.2765 &           0.3643 &           0.4517 &           0.5907 &           0.6735 \\
       & $xent_{or}+0.1*dice_{soft}$ &           0.0818 &           0.2353 &           0.3914 &           0.5056 &           0.6415 &           0.6860 &           0.2687 &  \textbf{0.3853} &           0.4391 &           0.5193 &           0.6238 &           0.6869 \\
       & $xent_{or}+dice_{log}$ &  \textbf{0.1883} &           0.3285 &           0.4417 &           0.5347 &           0.5874 &           0.6419 &              - &              - &              - &              - &              - &              - \\
       & $xent_{plus}+0.1*dice_{soft}$ &           0.0801 &           0.2085 &  \textbf{0.4547} &           0.5718 &  \textbf{0.6419} &  \textbf{0.6985} &              - &              - &              - &              - &              - &              - \\
       & $xent_{plus}+dice_{log}$ &           0.0900 &  \textbf{0.3356} &           0.3805 &  \textbf{0.5735} &           0.6236 &           0.6528 &              - &              - &              - &              - &              - &              - \\
\cline{1-14}
\multirow{8}{*}{spleen} & $xent_{base}$ &              - &              - &              - &              - &              - &              - &  \textbf{0.4295} &           0.4611 &           0.6961 &           0.7041 &           0.6663 &           0.8973 \\
       & $xent_{base}+0.1*dice_{soft}$ &              - &              - &              - &              - &              - &              - &           0.3496 &           0.5091 &           0.7621 &           0.6736 &           0.8338 &           0.8907 \\
       & $xent_{base}+dice_{log}$ &              - &              - &              - &              - &              - &              - &           0.0000 &           0.4490 &  \textbf{0.7941} &           0.6855 &           0.8011 &           0.8977 \\
       & $xent_{plus}$ &           0.5761 &           0.6657 &           0.7750 &  \textbf{0.8316} &           0.8531 &           0.8890 &           0.3746 &  \textbf{0.5632} &           0.6481 &           0.7809 &           0.7681 &           0.8506 \\
       & $xent_{or}+0.1*dice_{soft}$ &           0.5784 &           0.7034 &           0.7697 &           0.7807 &           0.8783 &           0.8605 &           0.3976 &           0.5624 &           0.7737 &  \textbf{0.7955} &  \textbf{0.8933} &  \textbf{0.9005} \\
       & $xent_{or}+dice_{log}$ &  \textbf{0.6639} &  \textbf{0.7318} &  \textbf{0.7993} &           0.7865 &           0.8351 &           0.8662 &              - &              - &              - &              - &              - &              - \\
       & $xent_{plus}+0.1*dice_{soft}$ &           0.5231 &           0.6780 &           0.7542 &           0.7583 &  \textbf{0.9044} &  \textbf{0.9009} &              - &              - &              - &              - &              - &              - \\
       & $xent_{plus}+dice_{log}$ &           0.6242 &           0.7182 &           0.7691 &           0.7694 &           0.8643 &           0.8421 &              - &              - &              - &              - &              - &              - \\
\bottomrule
\end{tabular}
}
	%\bigskip
	%\bigskip
	%\includegraphics[width=\linewidth,page=1]{MSD2018_smalldataset/seaborn_plot}
\end{table*}

\begin{table*}[htbp]%
    \caption{Dice performance of classifiers on the calf and thigh datasets. Dice scores in bold correspond to the best performing classifier in each label class. \textit{pl+edl} represents the pixel-wise union of \textit{peroneus longus} and \textit{extensor digitorum longus}. Performance of calf \gls*{imaf} could not be evaluated since it is not manually annotated in the calf datasets.}%
	\label{tab:musc_result}
	\centering%
	\resizebox{\linewidth}{!}{\begin{tabular}{lllllllllllllllllllll}
\toprule
label\_class & \multicolumn{2}{l}{\shortstack[l]{SCF}} & \multicolumn{2}{l}{\shortstack[l]{bone}} & \multicolumn{2}{l}{\shortstack[l]{vessels}} & \multicolumn{2}{l}{\shortstack[l]{tibialis\\ anterior}} & \multicolumn{2}{l}{\shortstack[l]{soleus}} & \multicolumn{2}{l}{\shortstack[l]{gastrocnemius}} & \multicolumn{2}{l}{\shortstack[l]{flexor\\ digitorum\\ longus}} & \multicolumn{2}{l}{\shortstack[l]{peroneus\\ longus}} & \multicolumn{2}{l}{\shortstack[l]{extensor\\ digitorum\\ longus}} & \multicolumn{2}{l}{\shortstack[l]{pl+edl}} \\
classifier\_type &          generic &         specific &          generic &         specific &          generic &         specific &           generic &         specific &          generic &         specific &          generic &         specific &                 generic &         specific &          generic &         specific &                   generic &         specific &          generic &         specific \\
loss\_function                 &                  &                  &                  &                  &                  &                  &                   &                  &                  &                  &                  &                  &                         &                  &                  &                  &                           &                  &                  &                  \\
\midrule
$dice_{soft}$                 &           0.8453 &           0.8523 &           0.9334 &           0.9448 &           0.4523 &           0.5820 &            0.8637 &           0.8787 &           0.8578 &  \textbf{0.9099} &           0.8483 &  \textbf{0.9146} &                  0.2431 &           0.8751 &           0.5247 &           0.8335 &                    0.6347 &           0.6231 &           0.6128 &  \textbf{0.8679} \\
$dice_{log}$                  &           0.8564 &           0.8608 &           0.9292 &           0.9338 &           0.5078 &           0.3816 &            0.8760 &           0.4279 &           0.8638 &           0.5314 &           0.8734 &           0.5241 &                  0.2752 &           0.6766 &           0.5068 &           0.2645 &                    0.4386 &           0.1175 &           0.5485 &           0.3619 \\
$xent_{base}$                 &           0.8214 &           0.7515 &           0.8129 &           0.9216 &           0.3430 &           0.4190 &            0.8159 &           0.8109 &           0.8755 &           0.8946 &           0.8769 &           0.9030 &                  0.8595 &           0.8841 &  \textbf{0.8349} &           0.8242 &                    0.6664 &           0.6656 &           0.8487 &           0.8557 \\
$xent_{base}+0.1*dice_{soft}$ &           0.8331 &           0.8699 &           0.9354 &           0.9247 &           0.4561 &           0.5822 &            0.8455 &           0.8748 &           0.8628 &           0.9068 &           0.8883 &           0.9117 &                  0.8706 &           0.8934 &           0.7446 &           0.8223 &                    0.5037 &           0.6415 &           0.8140 &           0.8539 \\
$xent_{base}+dice_{log}$      &           0.8585 &           0.8676 &           0.9312 &           0.9397 &           0.5439 &  \textbf{0.5865} &            0.8702 &           0.8800 &           0.8858 &           0.8903 &  \textbf{0.9045} &           0.8948 &                  0.8926 &           0.8282 &           0.8201 &           0.8177 &                    0.6129 &           0.6107 &           0.8296 &           0.8415 \\
$xent_{or}$                   &           0.8610 &           0.8610 &           0.9313 &           0.9474 &           0.5017 &           0.5171 &            0.8670 &           0.8914 &           0.8844 &           0.8936 &           0.8841 &           0.8996 &                  0.8844 &           0.8865 &           0.8080 &           0.7968 &                    0.6320 &           0.6637 &           0.8416 &           0.8378 \\
$xent_{or}+0.1*dice_{soft}$   &           0.8331 &           0.8663 &           0.9393 &           0.9480 &           0.5302 &           0.5572 &            0.8319 &           0.8942 &  \textbf{0.9050} &           0.9007 &           0.8715 &           0.8937 &                  0.8762 &           0.8864 &           0.7226 &           0.7878 &                    0.5800 &           0.6295 &           0.7834 &           0.8279 \\
$xent_{or}+dice_{log}$        &           0.8555 &  \textbf{0.8710} &           0.9228 &           0.9455 &           0.5336 &           0.5672 &   \textbf{0.9029} &           0.8004 &           0.8829 &           0.8426 &           0.8762 &           0.8328 &                  0.8869 &           0.8488 &           0.7881 &           0.7817 &           \textbf{0.7057} &           0.6117 &           0.8412 &           0.8224 \\
$xent_{plus}$                 &           0.8519 &           0.8555 &           0.9250 &  \textbf{0.9482} &           0.4809 &           0.5183 &            0.8411 &           0.8830 &           0.8601 &           0.8982 &           0.8686 &           0.8999 &                  0.8610 &           0.8978 &           0.7230 &           0.8095 &                    0.6292 &           0.6366 &           0.7605 &           0.8511 \\
$xent_{plus}+0.1*dice_{soft}$ &           0.8324 &           0.8607 &           0.9392 &           0.9435 &           0.5231 &           0.5639 &            0.8136 &  \textbf{0.8954} &           0.8896 &           0.8968 &           0.8601 &           0.8957 &                  0.8729 &  \textbf{0.9034} &           0.7029 &           0.7858 &                    0.5793 &  \textbf{0.6663} &           0.7783 &           0.8328 \\
$xent_{plus}+dice_{log}$      &  \textbf{0.8709} &           0.8646 &  \textbf{0.9418} &           0.9472 &  \textbf{0.5655} &           0.5695 &            0.8332 &           0.8735 &           0.8928 &           0.8983 &           0.9023 &           0.8875 &         \textbf{0.9026} &           0.8834 &           0.8151 &  \textbf{0.8343} &                    0.6468 &           0.6133 &  \textbf{0.8635} &           0.8640 \\
\bottomrule
\end{tabular}
}
	\resizebox{\linewidth}{!}{\begin{tabular}{llllllllllllllll}
\toprule
label\_class &             \shortstack[l]{IMAF} &              \shortstack[l]{SCF} &             \shortstack[l]{bone} &          \shortstack[l]{vessels} &  \shortstack[l]{adductor\\ magnus} & \shortstack[l]{biceps\\ femoris\\ long} & \shortstack[l]{biceps\\ femoris\\ short} &         \shortstack[l]{gracilis} &   \shortstack[l]{rectus\\ femoris} &        \shortstack[l]{sartorius} & \shortstack[l]{semimem-\\ branosus} &  \shortstack[l]{semiten-\\ dinosus} & \shortstack[l]{vastus\\ intermedius} & \shortstack[l]{vastus\\ lateralis} &  \shortstack[l]{vastus\\ medialis} \\
classifier\_type &         specific &         specific &         specific &         specific &         specific &            specific &             specific &         specific &         specific &         specific &         specific &         specific &           specific &         specific &         specific \\
loss\_function                 &                  &                  &                  &                  &                  &                     &                      &                  &                  &                  &                  &                  &                    &                  &                  \\
\midrule
$dice_{soft}$                 &  \textbf{0.7344} &           0.9453 &           0.9689 &  \textbf{0.6232} &           0.8325 &              0.8786 &      \textbf{0.8279} &           0.8254 &           0.0000 &           0.0000 &           0.9014 &           0.0000 &             0.5745 &           0.0000 &           0.6033 \\
$dice_{log}$                  &           0.7253 &           0.9377 &           0.9674 &           0.4789 &           0.7948 &              0.8429 &               0.7560 &           0.8063 &           0.7131 &           0.8702 &           0.5873 &           0.6815 &             0.8156 &           0.5190 &           0.8387 \\
$xent_{base}$                 &           0.7015 &           0.9464 &  \textbf{0.9728} &           0.5712 &  \textbf{0.8617} &     \textbf{0.8971} &               0.8200 &  \textbf{0.8656} &           0.8278 &           0.8997 &  \textbf{0.9130} &           0.8031 &             0.8515 &           0.8895 &           0.8723 \\
$xent_{base}+0.1*dice_{soft}$ &           0.7141 &           0.9432 &           0.9726 &           0.5513 &           0.8375 &              0.8733 &               0.7843 &           0.8190 &           0.8122 &           0.8864 &           0.8994 &           0.7811 &             0.8360 &  \textbf{0.8966} &           0.8911 \\
$xent_{base}+dice_{log}$      &           0.7048 &           0.9350 &           0.9590 &           0.4812 &           0.8402 &              0.8734 &               0.7831 &           0.8389 &           0.6542 &           0.8873 &           0.8812 &           0.7929 &             0.8347 &           0.8435 &           0.8638 \\
$xent_{or}$                   &           0.6911 &           0.9405 &           0.9658 &           0.5220 &           0.8414 &              0.8546 &               0.7799 &           0.8287 &           0.8089 &           0.8523 &           0.8916 &           0.7658 &             0.8402 &           0.8794 &           0.8613 \\
$xent_{or}+0.1*dice_{soft}$   &           0.7154 &           0.9433 &           0.9721 &           0.5766 &           0.8419 &              0.8657 &               0.8014 &           0.8594 &  \textbf{0.8373} &           0.8807 &           0.8869 &           0.7888 &             0.8519 &           0.8910 &           0.8987 \\
$xent_{or}+dice_{log}$        &           0.7248 &           0.9445 &           0.9718 &           0.5631 &           0.8594 &              0.8738 &               0.7846 &           0.8311 &           0.7540 &  \textbf{0.9008} &           0.9118 &  \textbf{0.8218} &             0.8274 &           0.8456 &           0.8672 \\
$xent_{plus}$                 &           0.6846 &  \textbf{0.9480} &           0.9317 &           0.4255 &           0.8247 &              0.8479 &               0.7350 &           0.7429 &           0.6899 &           0.8661 &           0.8900 &           0.7484 &             0.8111 &           0.8695 &           0.8488 \\
$xent_{plus}+0.1*dice_{soft}$ &           0.7107 &           0.9406 &           0.9641 &           0.6142 &           0.8476 &              0.8073 &               0.8142 &           0.8175 &           0.8227 &           0.8822 &           0.8946 &           0.7840 &    \textbf{0.8604} &           0.8941 &  \textbf{0.9000} \\
$xent_{plus}+dice_{log}$      &           0.7118 &           0.9406 &           0.9635 &           0.4904 &           0.8141 &              0.8843 &               0.7885 &           0.7646 &           0.6694 &           0.8831 &           0.8915 &           0.8184 &             0.8365 &           0.8391 &           0.8668 \\
\bottomrule
\end{tabular}
}
\end{table*}

    \begin{figure*}[!p]
		\begin{subfigure}[t]{\linewidth}
			\centering
			\includegraphics[width=\linewidth,page=2]{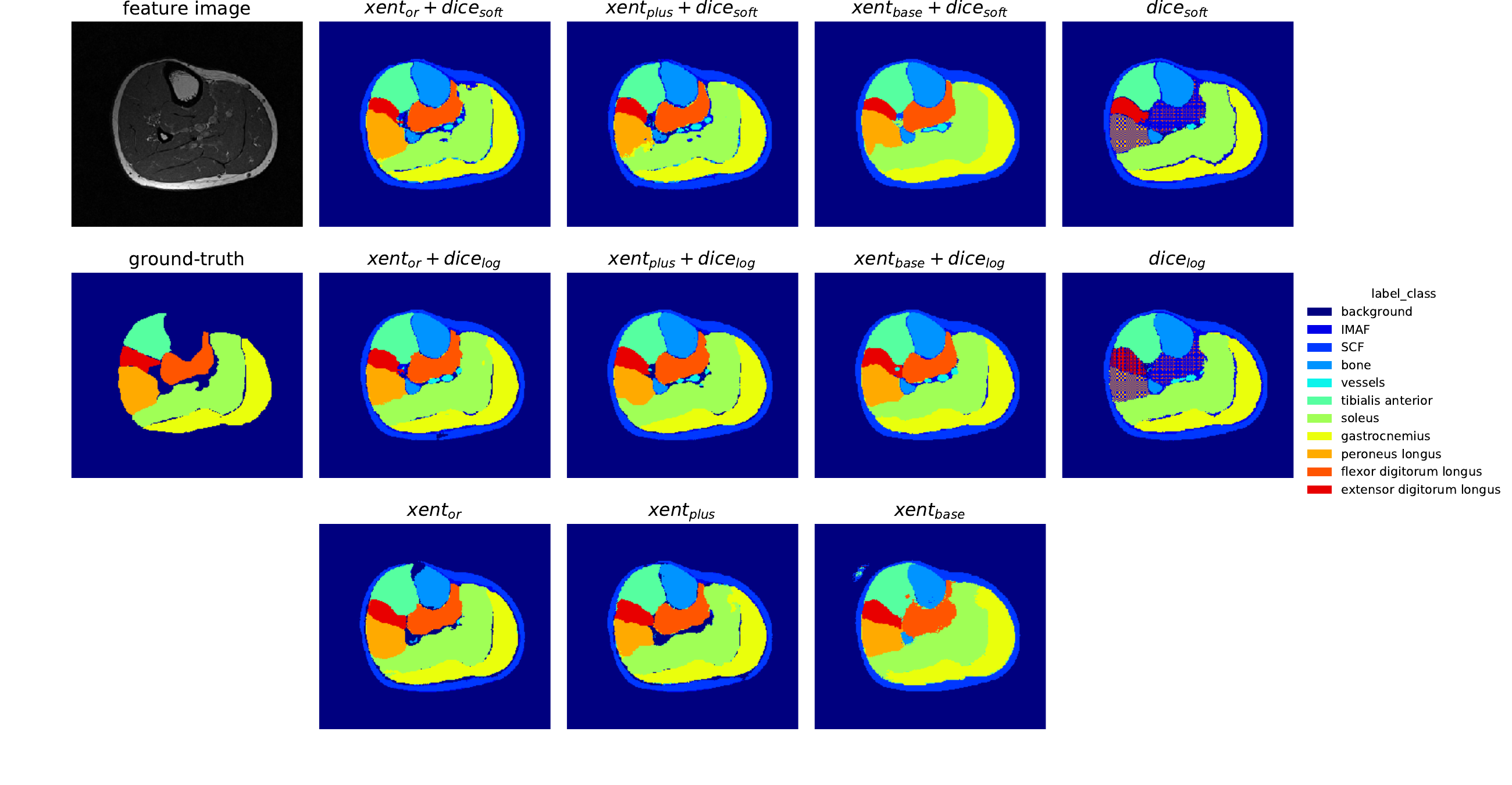}
			\caption{}%\caption{Predicted calf label map}
			\label{fig:musc_screenshot_calf}
		\end{subfigure}
		\begin{subfigure}[t]{\linewidth}
			\centering
			\includegraphics[width=\linewidth,page=2]{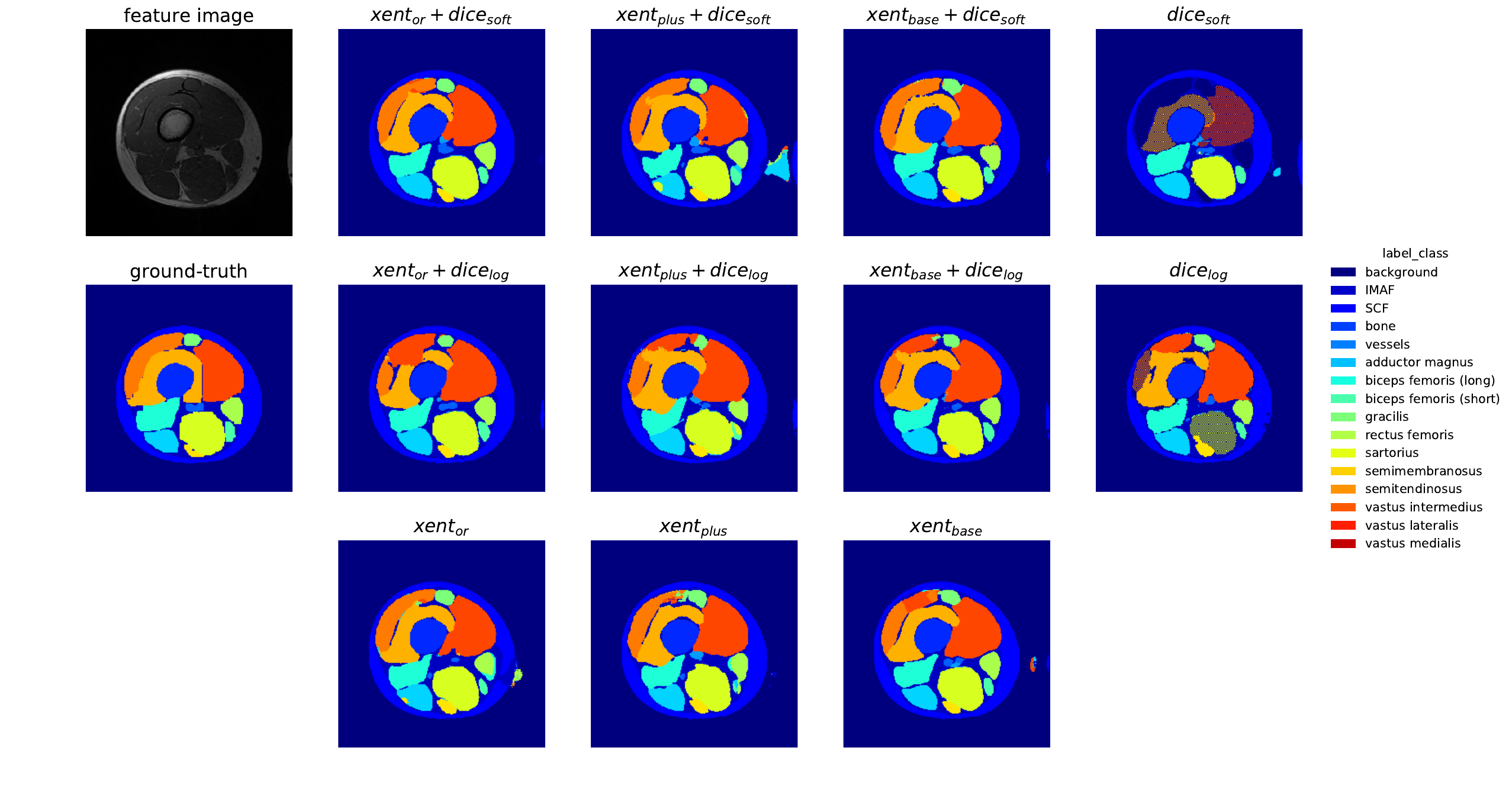}
			\caption{}%\caption{Predicted thigh label map}
			\label{fig:musc_screenshot_thigh}
		\end{subfigure}
		\caption{Label maps predicted by (\protect\subref{fig:musc_screenshot_calf}) calf generic-classifiers and (\protect\subref{fig:musc_screenshot_thigh}) thigh specific-classifiers.}
		\label{fig:musc_screenshot}
	\end{figure*}
%}

\end{document}